\global\def\draftcontrol{0}

   \def\versionno{Fragmenting Calogero -- draft   }

\catcode`\@=11

\expandafter\ifx\csname draftcontrol\endcsname\relax\global\def\draftcontrol{0}
\fi

{\count255=\time\divide\count255 by 60
\xdef\hourmin{\number\count255}
\multiply\count255 by-60\advance\count255 by\time
\xdef\hourmin{\hourmin:\ifnum\count255<10 0\fi\the\count255}}
\def\draftdate{\number\month/\number\day/\number\year\ \ \ \hourmin }

\newcommand\makepapertitle{\par
  \begingroup
    \renewcommand\thefootnote{\@fnsymbol\c@footnote}%
    \def\@makefnmark{\rlap{\@textsuperscript{\normalfont\@thefnmark}}}%
    \long\def\@makefntext##1{\parindent 1em\noindent
            \hb@xt@1.8em{%
                \hss\@textsuperscript{\normalfont\@thefnmark}}##1}%
     \newpage
     \global\@topnum\z@   
     \@makepapertitle
     \thispagestyle{empty}\@thanks
  \endgroup
  \setcounter{footnote}{0}%
  \global\let\thanks\relax
  \global\let\makepapertitle\relax
  \global\let\@makepapertitle\relax
  \global\let\@thanks\@empty
  \global\let\@author\@empty
  \global\let\@date\@empty
  \global\let\@title\@empty
  \global\let\title\relax
  \global\let\author\relax
  \global\let\date\relax
  \global\let\and\relax
  \def\version{\let\version\@version\@gobble}
}
\def\@makepapertitle{%
  \newpage
   \ifnum\draftcontrol=1 {}
   \version\versionno
   \vskip 3em%
   \else
   \hfill\hbox to 3cm {\parbox{4cm}{\@pubnum}\hss}%
   \vskip 3em%
   \fi
   \begin{center}%
   \let \footnote \thanks
     {\LARGE {\@title}}%
     \vskip 1.5em%
     {\normalsize
       \lineskip .5em%
       \begin{tabular}[t]{c}%
         \@author
       \end{tabular}\par}%
     \vskip 1.5em%
     {\@bstract}%
     \end{center}%
     \vskip 1.5em
     \@date%
   \par
}

\gdef\@pubnum{}
\def\pubnum#1{%
  \gdef\@pubnum{#1}}

\gdef\@bstract{}
\def\Abstract#1{%
  \gdef\@bstract{%
   \parbox{\textwidth-0pc}{%
   \centerline{\bf Abstract}\penalty1000%
\kern.2cm%
\noindent
\renewcommand\baselinestretch{1.0}%
{#1}}}
}

\def\ps@paper{\let\@mkboth\@gobbletwo%
     \ifnum\draftcontrol=1
    \def\@oddfoot{\hbox to \textwidth{\tiny \versionno \hfil\tiny\draftdate}%
    \hskip -\textwidth \hbox to \textwidth{\hfil\rm\thepage\hfil}}%
     \else\def\@oddfoot{\hbox to \textwidth{\hfil\rm\thepage\hfil}}
     \fi
     \let\@evenfoot\@oddfoot
}

\def\body{\clearpage
          \pagestyle{paper}
    }

\def\@version#1{\ifnum\draftcontrol=1
\typeout{}\typeout{#1}\typeout{}
\vskip3mm\centerline{\hbox{\fbox{\normalsize{\tt DRAFT -- #1 -- }
                   {\draftdate}}}}\vskip3mm
\fi}
\let\version\@version
\long\def\eqlabel#1{\ifnum\draftcontrol=1
                    \tag@false  
                    \tag*{(\theequation) \hbox to -0.2cm{\hspace{0cm}\small{#1}\hss}}
                    \refstepcounter{equation}
                    \edef\@currentlabel{\theequation}
                    \ltx@label{#1}          
                    \else
                    \label{#1}
                    \fi
                    }
\let\st@bibitem\@bibitem
\let\st@lbibitem\@lbibitem
\ifnum\draftcontrol=1
  \def\@bibitem#1{%
    \st@bibitem{#1}\a@@label{#1}\ignorespaces}
  \def\@lbibitem[#1]#2{%
    \st@lbibitem[#1]{#2}\a@@label{#2}\ignorespaces}
  \def\a@@label#1{%
    \gdef\a@lab{\smash{\normalfont\small#1}}
    \ifvmode
      \if@inlabel
        \global\setbox\@labels\hbox{%
          \llap{\a@lab\let\a@lab\relax
                \kern\@totalleftmargin\kern\marginparsep}%
          \box\@labels}%
      \fi
    \fi}
\fi

\documentclass[11pt,A4paper]{article}

\usepackage{amsmath,amssymb,array,calc,epsfig}
\usepackage{cite}
\usepackage[
      colorlinks=true,
      linkcolor=red,  
      urlcolor=blue,    
      filecolor=red,     
      linkcolor=blue,
      citecolor = red,
      pdfstartview=FitV,
      pdftitle={},
        pdfauthor={Paolo Benincasa},
        pdfsubject={},
        pdfkeywords={},
        pdfpagemode=None,
        bookmarksopen=true    
      ]{hyperref}
      
\usepackage{graphicx}
\usepackage{epstopdf}
\usepackage{ccaption}
\usepackage[utf8]{inputenc}

\ifnum\draftcontrol=1
\fi

\renewcommand\baselinestretch{1.25}
\renewcommand\section{\@startsection {section}{1}{\z@}%
                                   {-3.5ex \@plus -1ex \@minus -.2ex}%
                                   {2.3ex \@plus.2ex}%
                                   {\normalfont\large\bfseries}}
\renewcommand\subsection{\@startsection{subsection}{2}{\z@}%
                                   {-3.25ex\@plus -1ex \@minus -.2ex}%
                                   {1.5ex \@plus .2ex}%
                                   {\normalfont\normalsize\bfseries}}
\renewcommand\subsubsection{\@startsection{subsubsection}{3}{\z@}%
                                   {-3.25ex\@plus -1ex \@minus -.2ex}%
                                   {1.5ex \@plus .2ex}%
                                   {\normalfont\normalsize\it}}
\renewcommand\paragraph{\@startsection{paragraph}{4}{\z@}%
                                   {-3.25ex\@plus -1ex \@minus -.2ex}%
                                   {1.5ex \@plus .2ex}%
                                   {\normalfont\normalsize\bf}}

\numberwithin{equation}{section}

\def\revise#1       {\raisebox{-0em}{\rule{3pt}{1em}}%
                     \marginpar{\raisebox{.5em}{\vrule width3pt\
                     \vrule width0pt height 0pt depth0.5em
                     \hbox to 0cm{\hspace{0cm}{%
                     \parbox[t]{4em}{\raggedright\footnotesize{#1}}}\hss}}}}

\def\sqr#1#2{{\vcenter{\vbox{\hrule height.#2pt
 \hbox{\vrule width.#2pt height#1pt \kern#1pt
 \vrule width.#2pt}\hrule height.#2pt}}}}

\def\aa1{\phi}
\def\cc1{\psi}

\catcode`\@=12

\begin{document}


\title{\bf An SLE approach to four dimensional black hole microstate entropy}

\pubnum{%
arXiv:1701.xxxx}
\date{January 2017}

\author{
\scshape Paolo Benincasa${}^{\dagger}$, Suresh Nampuri${}^{\ddagger}$\\[0.4cm]
\ttfamily ${}^{\dagger}$Instituto de F{\'i}sica Te{\'o}rica, \\
\ttfamily Univerisdad Aut{\'o}noma de Madrid / CSIC \\
\ttfamily Calle Nicolas Cabrera 13, Cantoblanco 28049, Madrid, Spain\\
\small \ttfamily pablowellinhouse@anche.no \\[0.2cm]
\ttfamily ${}^{\ddagger}$CAMGSD-IST,\\
\ttfamily Universidade de Lisboa,\\
\ttfamily Av. Rovisco Pais, 1049-001, Lisbon, Portugal\\
\small \ttfamily nampuri@gmail.com
}

\Abstract{In this note, we model the  Bekenstein-Hawking entropy of a four dimensional extremal black hole in terms of  classifying particles moving in its  near horizon AdS$_2$ geometry. We use the framework of SLE curves in AdS$_2$ to classify these particle 
trajectories in terms of their boundary conditions. These turn out to be related to singular vectors in two-dimensional conformal quantum gravity theory in AdS$_2$ and the dynamics of these particles are governed by the Hamiltonians
of the integrable Calogero-like models, for these boundary conditions. We use this classification to count the leading order Bekenstein-Hawking entropy of the black hole and arrive at a first principle microscopic computation of black 
hole degeneracy.}

\makepapertitle

\body

\version\versionno

\section{Introduction}\label{sec:Intro}

The presence of a two-dimensional Anti-de-Sitter (AdS$_2$) space-time factor is a universal feature of extremal near-horizon black hole geometries. The near-horizon backgrounds are themselves solutions of the equations of motion of the
supergravity theory which contains the extremal black hole solutions, and in such theories, the neutral vector multiplet scalars flow from their asymptotic values to the extremum values of the black hole potential governing the flow, at 
the horizon, unless they constitute flat directions for the potential. Hence, the near-horizon region acts as an attractor for the flowing scalar moduli \cite{Ferrara:1996dd,Goldstein:2005hq}. Consequently, in the near horizon geometry,
they take values completely determined by the charges of the black hole, and hence the black hole horizon is characterized purely in terms of its charges so that the Bekenstein-Hawking entropy of the black hole is given in terms of a 
U-duality invariant composition of the charges. 

From the standpoint of quantum gravity,  an inevitable approach to formulating  black hole entropy from a statistical point of view is to explore the holographically dual conformal field theory (CFT). This approach has met with 
prominent success in the case of five-dimensional extremal black holes whose near-horizon geometry is a BTZ black hole in AdS$_3$ and consequently can be interpreted as a thermal ensemble in the dual CFT$_2$ \cite{Aharony:1999ti}. 
Then the high temperature expansion of the entropy of the ensemble in the CFT is independent of the couplings in the theory, has a universal dependence on the central charge and dilatation eigenvalue of the ensemble and is given by the 
Hardy-Cardy-Ramanujan formula \cite{Vafa:1995bm}. This accurately reproduces the Bekenstein-Hawking entropy. In the case of four-dimensional extremal black holes, one would need a precise formulation of the dual CFT$_1$, which has 
hitherto been unclear\footnote{Recently, there has been extensive work on this respect either by considering the two-dimensional theory as coming from a reduction of three-dimensional gravity with negative cosmological constant 
\cite{Cvetic:2016eiv} or analysing the SYK model of Majorana fermions with random few-point interactions \cite{Kitaev:2015kup, Maldacena:2016hyu} as a candidate for the holographic dual \cite{Maldacena:2016upp, Engelsoy:2016xyb} of 
particular dilaton-gravity systems \cite{Teitelboim:1983ux, Jackiw:1984je, Almheiri:2014cka}.}.
 
One model for a coarse understanding of charged extremal black hole statistical mechanics emerging from AdS$_2$ is motivated by the observation that under a KK uplift to five 
dimensions, the near horizon geometry yields a BTZ black hole in AdS$_3$, where the magnetic charges determine the AdS radius and the central charge of the dual CFT$_2$, while the electric charges determine the excitation number  above 
the vacuum \cite{Nampuri:2007gw} in terms of $L_0$, the dilatation eigenvalue of the dual chiral thermal ensemble\footnote{As an illustrative example, in  terms of the IIA ${\cal N}=2, D= 4$, a bound state of three types of D$4$ branes,
($p^a,\:a \,\in\, \{1,2,3\}$) and D$0$ ($q_0$) branes with triple intersection number $d_{abc}$ forms the microscopic weak coupling picture of an extremal black hole, with R$_{AdS}= d_{abc}= 6 d_{abc}p^a p^b p^c$ and $L_0= q_0$.}

In a similar fashion, we consider extremal particles (of unit specific charge) moving in AdS$_2$, and compute the dimension of the Hilbert space of this multi-particle system in AdS$_2$, viewed as a unit disk, by tracking the particle 
geodesics and classifying  these geodesics by first identifying them with the Schramm(stochastic)-L{\"o}wner evolution (SLE) \cite{Loewner:1923se, Schramm:1999rp} curves\footnote{For a review about the SLE see \cite{Cardy:2005kh}.} and 
hence delineating their boundary conditions on the unit circle. Thereby, we compute the dimension of the phase space of these geodesics to show that it determines the black hole entropy, and indicate how this naturally leads to 
fragmented AdS$_2$ geometries \cite{Maldacena:1998uz} being candidates for microstate backgrounds for the extremal black hole.

Interestingly, this system is described in terms of the Calogero-Sutherland (CS) model \cite{Calogero:1969xj, Calogero:1969ie, Calogero:1970nt, Sutherland:1971kq, Sutherland:1971ks, Calogero:1975ii, Sutherland:1975er}, which is a fully 
integrable quantum mechanical $N$-particle system on a circle with a pairwise long range interaction\footnote{For more details see the review \cite{Polychronakos:2006nz} and references therein.}, and thus the analysis of particle 
trajectories  can be cast in terms of its Hamiltonians and their eigenfunctions. These, in turn,  are directly related to a specific class of correlators in the boundary conformal field theory (BCFT) of quantum gravity on AdS${}_2$,  
viewed as a unit disk.

The paper is organised as follows. In Section \ref{sec:BCFT}, we give a short elucidation of the features of BCFT relevant for our discussion as well as briefly enumerate the Calogero-Sutherland integral invariants. We also outline the
correlation between generators of the BCFT and the Calogero- Sutherland eigenfunctions. We follow this up with presenting our calculational methodology for analyzing particle trajectories in AdS$_2$, in Section \ref{sec:Meth} and 
establish our basic working computations that cast the particle classification problem in terms of boundary operators in the BCFT. We present general results based on this connection in Section \ref{sec:BCFT-CS} and finally, use it to 
compute the Bekenstein-Hawking entropy in Section \ref{sec:BH}. We finally end with a discussion of future possibilities and open questions.


\section{Boundary CFTs in Two Dimensions}\label{sec:BCFT}

Quantum gravity in global AdS${}_2$ can be thought of as a two-dimensional boundary conformal field theory (BCFT${}_2$) with states given by degenerate boundary primary fields (the so-called  ZZ boundary 
primaries) \cite{Zamolodchikov:2001ah}. It can be viewed as the standard two-dimensional Liouville theory \cite{Seiberg:1990eb, Nakayama:2004vk} (with ZZ boundary conditions) with appropriate matter thrown in to nullify the central charge. In general, and unless specified, 
we focus on the gravity sector only, which corresponds to a BCFT${}_2$ with a central charge which can be parametrised as 
\begin{equation}\eqlabel{eq:BCFTcc}
 c\:=\:1-6\frac{(g-1)^2}{g},
\end{equation}
with $g\,\in\,\mathbb{C}\,\backslash\,\{0\}$. This form of the central charge is possible if degenerate states exist. In order to have a self-contained exposition, in this section we will discuss the salient features of BCFT${}_2$.

On a manifold with boundaries the boundary conditions break the algebra to be just one copy of the Virasoro algebra and the set of possible boundary conditions
is isomorphic to the space of conformal blocks \cite{Cardy:1989ir}.
 
Let us begin with considering a CFT${}_2$ on an infinitely-long strip parametrised by the coordinates $(\tau,\,\sigma)$ and with the two boundaries at $\sigma\,=\,0,\,\pi$. On each of the boundary, it is possible to choose different 
boundary conditions: The general requirement which ensures that conformality is preserved at $\sigma\,=\,0,\,\pi$ is
the vanishing of the off-diagonal components of the stress tensor, $T_{\tau\sigma}\,=\,0$, but otherwise
one can pick an arbitrary pair $(\alpha,\,\beta)$ of boundary conditions on $\sigma\,=\,(0,\,\pi)$ which
is consistent with the above condition. The Hamiltonian $H_{\alpha\beta}$, {\it i.e.} the generator of the
translations in $\tau$ with the boundary conditions $(\alpha,\,\beta)$, has its eigenstates in the irreducible
representations of the (single) Virasoro algebra.

The infinitely-long strip can be mapped into the upper half plane via $z\,=\,e^{\tau+i\sigma}$
\begin{equation*}
 \raisebox{-2cm}{\scalebox{.40}{\includegraphics{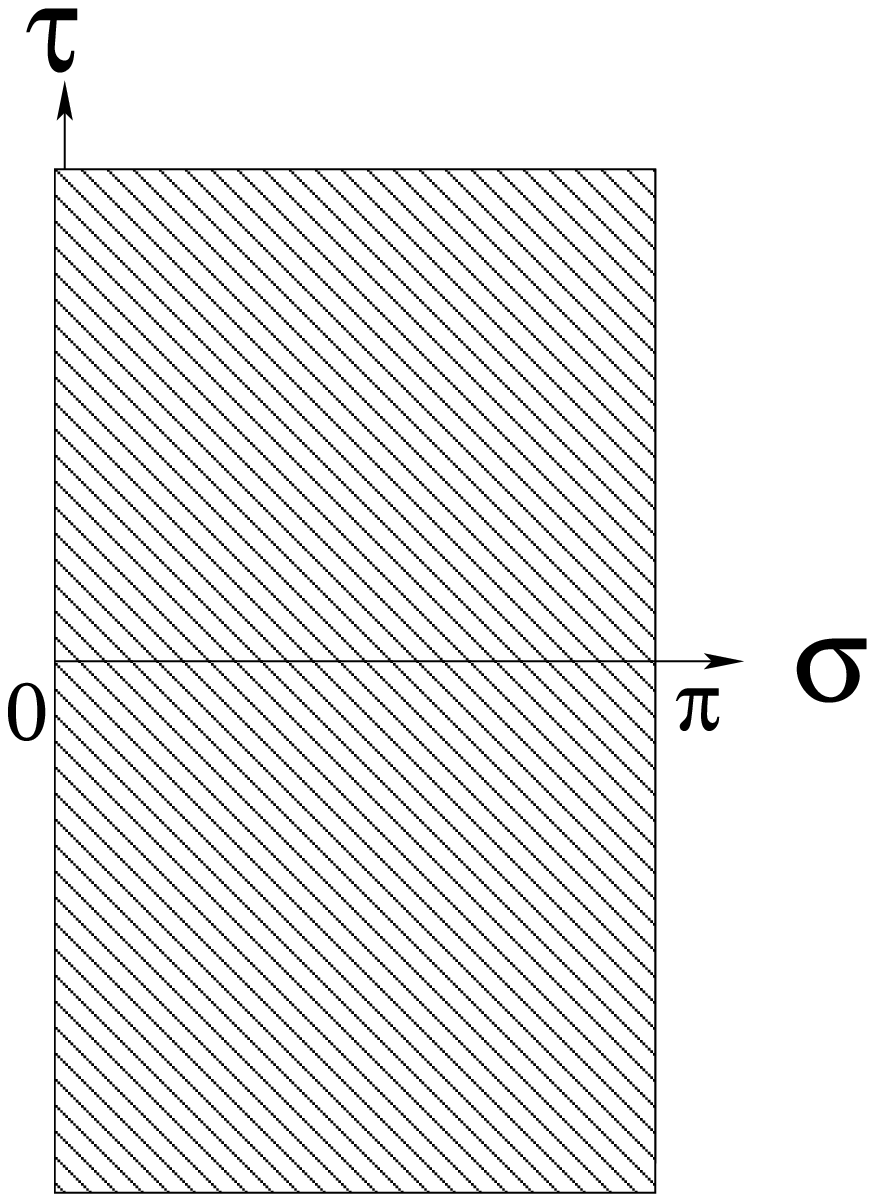}}}
 \qquad
 \xrightarrow{z\:=\:e^{\tau+i\sigma}}
 \qquad
 \raisebox{-1.8cm}{\scalebox{.50}{\includegraphics{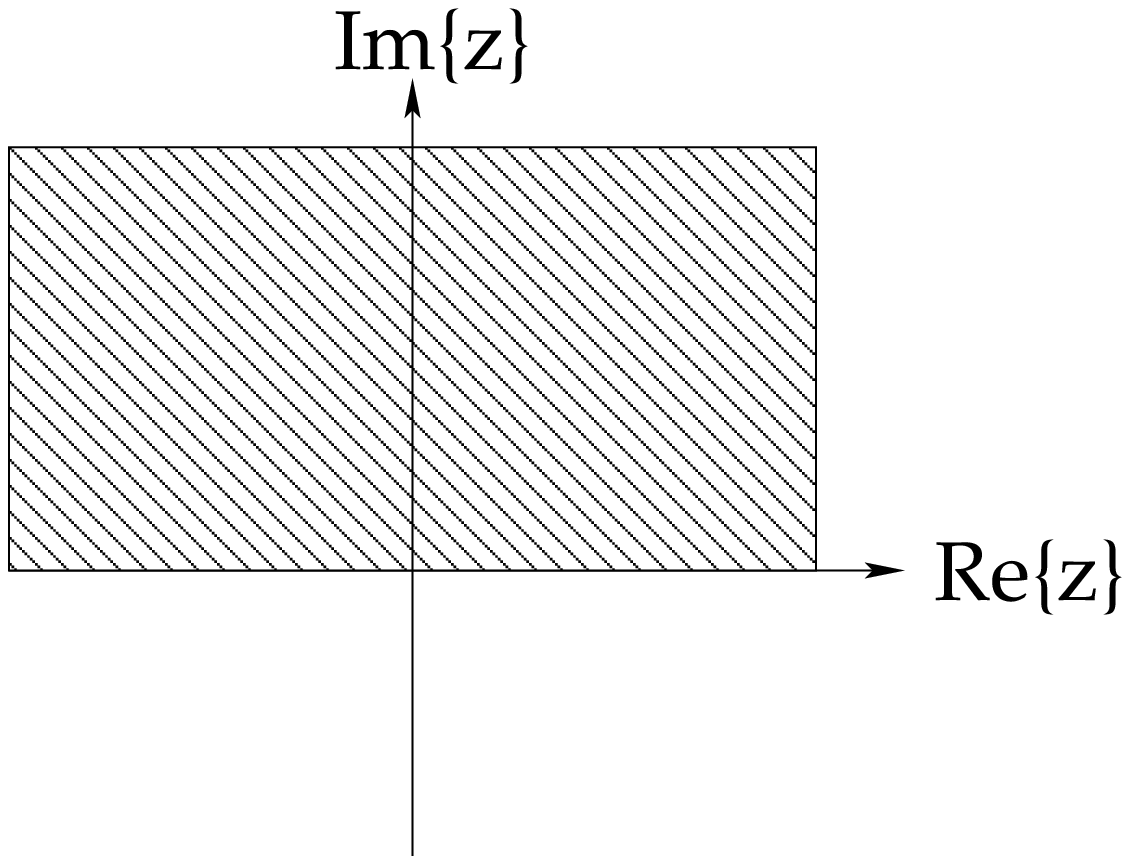}}}
\end{equation*}
where the two boundaries are merged into a single one which coincides with the real axis. In this case
the class of conformal boundary conditions is given by the requirement that the stress tensor is real
on the real axis ($T(z)\,=\,\bar{T}(\bar{z})$). The Virasoro generators are given by
\begin{equation}\eqlabel{eq:VirGen}
 L_n\:=\:\frac{1}{2\pi i}\int_{\mathcal{C}_{+}}dz\,z^{n+1}T(z)-\frac{1}{2\pi i}\int_{\mathcal{C}_{+}}d\bar{z}\,\bar{z}^{n+1}\bar{T}(\bar{z}),
\end{equation}
with $\mathcal{C}_{+}$ being a semi-circle in the upper half plane (UHP). Because of the reality condition of the
stress tensor on the real line, $\bar{T}(\bar{z})$ can be regarded as the analytic continuation of $T(z)$ in the
lower half plane and, consequently, the Virasoro generators can be expressed as a single integral of the stress tensor
along a full circle $\mathcal{C}$:
\begin{equation}\eqlabel{eq:VirGen2}
 L_n\:=\:\frac{1}{2\pi i}\oint_{\mathcal{C}}dz\,z^{n+1}T(z),
\end{equation}
The dilatation operator $L_0$ corresponds to the Hamiltonian $H$ of the infinitely-long stripe, up to a central charge dependent term:
\begin{equation}\eqlabel{eq:Dil}
 L_0\:=\:H + \frac{c}{24}.
\end{equation}
The eigenstates of $L_0$ correspond to primary boundary operators $\phi(0)$ which act on the vacuum. If in the infinitely-long stripe we consider
a pair $(\alpha,\,\beta)$ of boundary conditions, in the upper half plane this corresponds to primary boundary operators $\phi_{\alpha\beta}(0)$
which changes the boundary conditions between the positive and negative real axis.
For the purpose of this note, we will be treating AdS$_2$ as a unit disk, $D$,  and the transformation from the UHP to $D$ is accomplished by the inverse M{\"o}bius transformation:
\begin{equation}\eqlabel{eq:Mtrsf}
z\quad\longrightarrow\quad w = -i\,\frac{z+i}{z-i}
\end{equation}
This transforms the real boundary of the UHP to the unit circle. 
For an anomaly free conformal theory, the stress tensor transforms as $T_{zz} \rightarrow (\frac{\partial z}{\partial w})^2 T_{zz}$, thus resulting in the conformal boundary condition,  
\begin{equation}\label{eq:CBCst}
\frac{1}{\bar{z}^4} T_{zz}\left(\frac{1}{\bar{z}}\right)\:=\:\left.\bar{T}_{\bar{z}\bar{z}}\right|_{z \, \epsilon \, \partial D}\,.
\end{equation}

Hence, at the boundary, \eqref{eq:CBCst} relates  the anti-holomorphic Virasoro generators to the holomorphic ones
\begin{equation}\eqlabel{eq:VirGenRel}
 \bar{L}_{k}\:=\:(-1)^k\bar{z}_i^{2k}\sum_{l \ge 0}\bar{z}_i^l\frac{(2-k-l)_l}{l!}L_{k+l}
\end{equation}
allowing us to 
 find  the modes $L_{-k}$ as differential operators acting on
the boundary.

The presence of a boundary does not change the OPEs, which are local quantities and hence the holomorphic stress tensor action on the primary field is given by the conventional form,
\begin{equation}\eqlabel{eq:OPE}
 T(z)\phi(w)\:\sim\:\frac{\Delta_{\phi}}{(z-w)^2}\phi(w)+\frac{1}{z-w}\partial_w\phi(w)+\ldots\:=\:\sum_{n=0}^{\infty}(z-w)^{n-2}L_{-n}\phi(w)\,.
\end{equation}
as is the anti-holomorphic equivalent.
However,  as a consequence of the existence of just one Virasoro algebra, the conformal Ward identity becomes
\begin{equation}\eqlabel{eq:CWI}
 \langle T(z)\prod_{i=1}^N\phi(w_i,\,\bar{w}_i)\rangle\:=\:
 \sum_{i=1}^N
 \left[
  \frac{\Delta_i}{(z-w_i)^2}+\frac{1}{z-w_i}\partial_{w_i} + \frac{\bar{\Delta}_i}{(z-w_i)^2} + \frac{1}{z-w_i}\partial_{w_i}
 \right]
 \langle\prod_{i=1}^N\phi(w_i,\,\bar{w}_i)\rangle.
\end{equation}
The OPE \eqref{eq:OPE} defines a higher order operator $L_{-n}\phi$ of conformal dimension $\Delta+n$ and reiterated applications of $L_{-n_i}$
defines further higher order operators with conformal dimension $\Delta + \sum_i^k n_i$ whose correlators are related to $\phi$ by differential operators. The conformal boundary condition 
and the  Ward identity ensure that \eqref{eq:CWI}, the  correlator $\langle\prod_{i=1}^N\phi(w_i,\bar{w}_i)\rangle$ as a 
function of $(\{w_i\},\{\bar{w}_i\})$ satisfies  the same differential equation as the {\it bulk} correlator $\langle\prod_{i=1}^{2N}\phi(w_i,\bar{w}_i)\rangle$,
viewed as a function of $(\{w_i\})$ only \cite{Cardy:1984bb}:
\begin{equation}\eqlabel{eq:DiffEq}
 \begin{split}
  &\langle\left[\prod_{i=1}^k L_{-n_i}\phi(z)\right]\prod_{j=1}^N\phi(w_i)\rangle\:=\:\prod_{i=1}^k\mathcal{L}_{-n_i}\langle\phi(z)\prod_{j=1}^N\phi(w_j)\rangle\\
  &\mathcal{L}_{-n_i}\:=\:\sum_{j=1}^N
    \left[
     \frac{(1-n_i)\Delta_j}{(z-w_j)^{n_i}}-\frac{1}{(z-w_j)^{n_i-1}}\partial_{w_j}
    \right]
 \end{split}
\end{equation}
If at a given $L\,=\,\sum_i^k n_i$ the descendent of $\phi$ are degenerate, then the correlators involving $\phi$ satisfy a differential equation of the form
\begin{equation}\eqlabel{eq:NullVectCond}
 \mathcal{O}(z)\langle\phi(z)\bullet\rangle\:=\:0,
\end{equation}
$\mathcal{O}(z)$ being a linear combination of the operators $\mathcal{L}_{-n_i}$ which can appear at level $L$, while ``$\bullet$'' just indicates an arbitrary
set of operators in the correlator.

The operators $\phi$ which are degenerate at level $L$ are characterised by a conformal dimension which can be thought of as dependent on some function of the central charge and two integers $r$ and $s$ such that $L\,=\,rs$. If we 
parametrise the central charge as in \eqref{eq:BCFTcc}, which enjoys the invariance under $g\,\longrightarrow\,1/g$, then the conformal dimension of these fields 
$\phi^{(r,s)}$ \footnote{It is convenient to use the labels $(r,s)$ to make manifest that a field is degenerate.} acquires the form
\begin{equation}\eqlabel{eq:DegConfDim}
 \Delta_{(r,s)}\:=\:\frac{1}{4}\left[\frac{r^2-1}{g}+(s^2-1)g + 2(1-rs)\right].
\end{equation}
It is important to remark that singular vectors of level $L\,=\,rs$ exist if and only if the central charge $c$ and the highest conformal weight $\Delta_{\mbox{\tiny $(r,s)$}}$ have the form \eqref{eq:BCFTcc} and \eqref{eq:DegConfDim}
respectively.

For future reference, let us write explicitly the null vector conditions \eqref{eq:NullVectCond} at levels $L\,=\,1,\,2\mbox{ and }3$, whose general form is
\begin{equation}\eqlabel{eq:NVC123}
 \mathcal{O}^{(g)}_L(z)\phi^{(r,s)}\:=\:0\qquad\Longrightarrow\qquad\mathcal{O}^{(g)}_L(z)\langle\phi^{(r,s)}\bullet\rangle\:=\:0,
\end{equation}
where the dependence on the central charge through $g$ as well as on the pair of integers $(r,s)$ is explicit. Thus:
\begin{itemize}
 \item $L\,=\,1$. There is just one operator with conformal dimension $\Delta_{(1,1)}\,=\,0$, which is nothing but the identity operator $\phi^{(1,1)}\,=\,\mathbb{I}$:
       \begin{equation}\eqlabel{eq:NVC1}
        \mathcal{O}^{(g)}_1(z)\phi^{(1,1)}(z)\,\equiv\,L_{-1}\phi^{(1,1)}(z)\,\equiv\,\partial_{z}\phi^{(1,1)}(z)\:=\:0
       \end{equation}
 \item $L\,=\,2$. There are two operators at this level, $\phi^{(1,2)}$ and $\phi^{(2,1)}$, with respectively conformal dimensions $\Delta_{(1,2)}\,=\,(3g-2)/4$
       and $\Delta_{(2,1)}\,=\,(3-2g)/4g$, which satisfy the following differential equations
       \begin{equation}\eqlabel{eq:NVC2}
        \begin{split}
         0\:&=\:\mathcal{O}^{(g)}_{2}(z)\langle\phi^{(1,2)}(z)\bullet\rangle\:\equiv\:\left[L_{-1}^2-gL_{-2}\right]\langle\phi^{(1,2)}(z)\bullet\rangle\:\\ 
         &\phantom{\ldots}\\
         0\:&=\:\mathcal{O}^{(1/g)}_2(z)\langle\phi^{(2,1)}(z)\bullet\rangle\:\equiv\:\left[L_{-1}^2-\frac{1}{g}L_{-2}\right]\langle\phi^{(2,1)}(z)\bullet\rangle\:
        \end{split}
       \end{equation}
 \item $L\,=\,3$. As in the previous case, there are two degenerate operators, namely $\phi^{(1,3)}$ and $\phi^{(3,1)}$, with conformal dimension respectively
       $\Delta_{(1,3)}\,=\,2g-1$ and $\Delta_{(3,1)}\,=\,2/g-1$, whose correlators satisfy the following third-order differential equation:
       \begin{equation}\eqlabel{eq:NVC3}
        \begin{split}
         0\:&=\:\mathcal{O}^{(g)}_3(z)\langle\phi^{(1,3)}(z)\bullet\rangle\:\equiv\:\left[L_{-1}^3-4gL_{-2}L_{-1}-2g(1-2g)L_{-3}\right]\langle\phi^{(1,3)}(z)\bullet\rangle\:\\
         &\phantom{\ldots}\\
         0\:&=\:\mathcal{O}^{(1/g)}_3(z)\langle\phi^{(3,1)}(z)\bullet\rangle\:\equiv\:\left[L_{-1}^3-\frac{4}{g}L_{-2}L_{-1}-\frac{2}{g}(1-\frac{2}{g})L_{-3}\right]\langle\phi^{(3,1)}(z)\bullet\rangle\:
        \end{split} 
       \end{equation}       
\end{itemize}
Notice that the equations for $\phi^{(r,s)}$ ad $\phi^{(s,r)}$ are related to each other via the transformation $g\,\longrightarrow\,1/g$, which is the very same leaving the central charge invariant. Furthermore,
the semi-classical limit $c\,\longrightarrow\,\infty$ can be taken in two ways in terms of $g$ by sending it either to $0^{-}$ or to $-\infty$. 


\subsection{From null vectors to integrable models: A first look}

The Virasoro singular vectors discussed in the previous section turn out to be in one-to-one correspondence with the Jack symmetric functions $J_{\lambda}$ graded by rectangular Young diagrams \cite{Mimachi:1995my}, which are 
eigenfunctions of the Laplace-Beltrami operator,
\begin{equation}\eqlabel{eq:LBop}
 \tilde{H}\:=\:\sum_{i=1}^N\left(z_i\partial_{z_i}\right)^2+g\sum_{i<j}\frac{z_i+z_j}{z_i+z_j}\left(z_i\partial_{z_i}-z_j\partial_{z_j}\right),
\end{equation}
with $g$ being a coupling constant and $\lambda$ being a lexicographically ordered partition $\lambda\,=\,[\lambda_1,\,\ldots,\,\lambda_{N}]$. The eigenvalues of $\tilde{H}$ in \eqref{eq:LBop} corresponing to the Jack symmetric 
eigenfunctions $J_{\mbox{\tiny $\lambda$}}$ are 
\begin{equation}\eqlabel{eq:LBopEv}
 \Lambda_2(g)\:=\:\sum_{i=1}^N\lambda_i
  \left[
   \lambda_i+g\left(N+1-2i\right)
 \right].
\end{equation} 
The operator is related to the (second order) Calogero-Sutherland Hamiltonian $H_2^{\mbox{\tiny $(g)$}}$ via a similarity transformation, with such a tranformation involving the Calogero-Sutherland 
vacuum $N$-particle wave-function 
$\Psi_{N}$:
\begin{equation}\eqlabel{eq:LB-CS}
 \begin{split}
  &H^{(g)}_2\:=\:\sum_{i=1}^N\left(z_i\partial_{z_i}\right)^2-g(g-1)\sum_{i\neq j}\frac{z_iz_j}{(z_i-z_j)^2},\qquad\Psi_N(z)\:=\:\prod_{i<j}\left(z_i-z_j\right)^{g/2},\\
  &\tilde{H}\:=\: [\Psi_N(z)]^{-1} H^{(g)}_2\Psi_N(z)-\varepsilon_0,\qquad\varepsilon_{0}\:\equiv\:\mbox{zero-point energy}.
 \end{split}
\end{equation}
The Calogero-Sutherland model describes a quantum mechanical $N$-particle system on a circle with a pairwise long range interaction (inverse square potential). It turns out to be completely integrable and thus caracterised by an 
infinite set of (commuting) integrals of motion. For later convenience, let us list the very first of them:
\begin{equation}\eqlabel{eq:CSintmot}
 \begin{split}
  &H^{(g)}_1\:=\:\sum_{i=1}^N z_i\partial_{z_i},\\
  &H^{(g)}_2\:=\:\sum_{i=1}^N\left(z_i\partial_{z_i}\right)^2-g(g-1)\sum_{i\neq j}\frac{z_iz_j}{(z_i-z_j)^2},\\
  &H^{(g)}_3\:=\:\sum_{i=1}^N\left(z_i\partial_{z_i}\right)^3-\frac{3}{2}g(g-1)\sum_{i\neq j}\frac{z_i z_j}{(z_i-z_j)^2}\left(z_{i}\partial_{z_i}+z_{j}\partial_{z_j}\right).
 \end{split}
\end{equation}
In general, as is true for an integrable system, the integrals are generated by a  $N \times N$ Lax matrix, 
$L_{kl}= z_i \partial_i \delta_{k,l}\,+\, \sum_{i \neq j}\frac{z_i}{z_i-z_j}\,(\,1\,-\,\delta_{k,l}\,)$ from which the the $n$-th invariant $H^{(g)}_n$ can be reproduced as 
\begin{equation}
H^{(g)}_n = Tr\left\{L^n\right\}
\end{equation}
Because of the similarity transformation \eqref{eq:LB-CS} between $\tilde{H}$ and $H_2^{\mbox{\tiny $(g)$}}$\footnote{Notice that the zero-point $\varepsilon_0$ in \eqref{eq:LB-CS} comes from a similarity transformation on $H_1$. In
other words, if we consider the similarity transformation $\Psi_N\tilde{H}\Psi_N^{-1}$ on $\tilde{H}$ (which is the inverse of \eqref{eq:LB-CS}), we obtain a linear combination of the first- and second-order Calogero-Sutherland 
Hamiltonians $H_1$ and $H_2$.}, the  eigenfunctions of the second order Calogero-Sutherland Hamiltonian $H_2^{\mbox{\tiny $(g)$}}$ acquire the form $\Psi_N(z)J_{\lambda}(z)$, with $J_{\lambda}(z)$ being the Jack symmetric polynomials 
which describe the Calogero-Sutherland excited states.


\section{Methodology}\label{sec:Meth}

Let us now consider AdS${}_2$ as a BCFT${}_2$ on a unit disk $D$, which can be obtained from the upper-half plane via the conformal transformation \eqref{eq:Mtrsf}. The spectrum of this theory is described by AdS$_2$ and fragments of 
AdS$_2$ in the conformal gauge, which correspond to the ZZ boundary conditions on the unit circle \cite{Giombi:2008sr}. The $s$-th ZZ boundary condition corresponding to the $(s-1)$-th excitation in the spectrum is the primary state, 
$|\phi^{\mbox{\tiny $(1,s)$}}\rangle$, having a null state at the $s$-th level, with the identity operator, $\mathbb{I}\,\equiv\,|\phi^{\mbox{\tiny $(1,1)$}}\rangle$, corresponding to the vacuum. Solving this is equivalent to knowing 
all correlation functions of this theory. We consider the problem of tracing particles moving in the unit disk. In particular, a particle moving from a boundary point $z_1$ on the  unit circle towards a bulk point $(w,\bar{w})$ with an
operator insertion $V(w,\bar{w})$ and returning to another point $z_2$ on the unit circle, corresponds to boundary operator insertions at $z_1$ and $z_2$ on the boundary. Since the degenerate primaries form a basis for these operators, 
we are naturally led to a basis for these curves which consists of curves intersecting the boundary at points corresponding to the insertion of a degenerate primary. Hence, computing the probability 
$\mathcal{P}(\gamma_s(w,\bar{w});\:z_1,\,z_2)$ of such a curve $\gamma_s$, yields the correlator of the corresponding degenerate boundary primaries localized at the intersection points, given by:
\begin{equation}\eqlabel{eq:Prob}
\mathcal{P}(\gamma_s(w,\bar{w});\:z_1,\,z_2)\:=\:\langle\phi^{\mbox{\tiny $(1,s)$}}(z_1)|V\left(w,\bar{w}\right)|\phi^{\mbox{\tiny $(1,s)$}}(z_2)\rangle.
\end{equation}
If the bulk operator $V(w,\bar{w})$ is taken to be an identity operator, then this bulk boundary operator is essentially a boundary-boundary correlation function.  These curves have the following properties:
\begin{enumerate}
\item As the space time has an $SL\left(2, \mathbb{R}\right)$ isometry, the measure on the curves is preserved under a conformal transformation. 
\item Under the usual insertion of basis of states,$\{\mathcal{O}_i\}(w,\bar{w})$ in a correlation function, 
\begin{equation}\eqlabel{eq:OinsCorr}
\langle\phi^{\mbox{\tiny $(1,s)$}}|\phi^{\mbox{\tiny $(1,s)$}}\rangle = \sum_i\langle\phi^{\mbox{\tiny $(1,s)$}}|\mathcal{O}_i(w,\bar{w})\rangle\langle\mathcal{O}_i\left(w,\bar{w}\right)|\phi^{\mbox{\tiny $(1,s)$}}\rangle\,,
\end{equation} 
we see that the conditional probability to go from the point $(w,\bar{w})$ to the final point give the trajectory $A$ from the initial point to $(w,\bar{w})$ is the same as the probability of a trajectory from $(w,\bar{w})$ to the 
final point in the domain of the unit disk modulo $A$.
\end{enumerate}
The above two conditions ensure \cite{Cardy:2005kh} that curves of the type $\gamma_s$ are a combination of two radial SLE curves, one stretching from a boundary point $z_{\mbox{\tiny $\partial$}}$ to a bulk point and another from the 
bulk to a second boundary point. We will henceforth take this bulk point to be at the origin, without loss of generality, as one can always conformally transform back the origin to the given bulk point by a boundary-preserving map:
\begin{equation}\eqlabel{eq:Ct}
z\quad\longrightarrow\quad \frac{z+w}{\bar{w}z+1}.
\end{equation}
Therefore, given $\lambda$ to be the affine parameter on a given SLE curve, then one can track the evolution of the curve from the boundary to the bulk by first mapping the point on the curve $z(\lambda)$  to the boundary of the disk 
via a mapping $g_{\mbox{\tiny $z_{\partial}$}}:\:D\rightarrow \partial D$ and subsequently tracing the curve in the bulk by tracing the evolution of the map with respect to $\lambda$. The corresponding equation must map the boundary of 
the unit disk to itself modulo the intersection point of the curve with the boundary and leave the origin invariant and must be a meromorphic function, with poles only on the boundary. These criteria restricts the curves to be given by 
the L{\"o}wner equation \cite{Cardy:2005kh}
\begin{equation}\eqlabel{eq:SLE}
\partial_\lambda g_{\mbox{\tiny $z_{\partial}$}}\left(z(\lambda)\right)\:=\:-g_{\mbox{\tiny $z_1$}} \frac{g_{\mbox{\tiny $z_{\partial}$}} + z_{\partial}}{g_{\mbox{\tiny $z_{\partial}$}}-z_{\partial}}
\end{equation}
The above equation satisfying the L{\"o}wner criteria governs the evolution of the SLE curve in terms of dynamics of points on the boundary that are shadows to points on this curve under the map $g$. In order to see the implications for 
the correlation function under consideration (which is a composite of two such curves), we simply apply a conformal transformation based on \eqref{eq:SLE} on it. The result of this conformal transformation on the correlation function 
of degenerate primaries ({\it i.e.} taking the bulk insertion $V$ to be the identity) is given as, 
\begin{equation}\label{eq:q1}
\left(
 \int_{\partial D_{\mbox{\tiny in}}} dz \,\alpha(z)\,T(z) +  \int_{\partial D_{\mbox{\tiny out}}} d\bar{z} \,\overline{\alpha (z)}\,\bar{T}(\bar{z})
\right) \langle\phi^{\mbox{\tiny $(1,s)$}}(z_1)|\mathbb{I}(0)|\phi^{\mbox{\tiny $(1,s)$}}(z_2)\rangle\:=\:0\,.
\end{equation}
Here, $D_{\mbox{\tiny in}}$ and $D_{\mbox{\tiny out}}$ signify the integration  done on a curve infinitesimally close to the unit circle in the counterclockwise direction, inside and out of the unit disk respectively. 
\begin{equation}\label{eq:SLEct}
\alpha_{\mbox{\tiny $2$}}(z)= - z \sum^2_{i=1} \frac{z+z_i}{z-z_i}\,,
\end{equation}
where the boundary points $z_i$ correspond to the starting and ending positions of the SLE curve or the position of the boundary operators. Note that the conformal map $\alpha_{\mbox{\tiny $2$}}(z)$ designed to satisfy the L{\"o}wner 
criteria is only one of a class of meromorophic functions with poles on the boundary of the unit disk, and indexed by $L\,\in\,\mathbb{N}^+$. A generic member of this L{\"o}wner function class is given by 
\begin{equation}\eqlabel{eq:aL}
\alpha_{\mbox{\tiny $L$}} (z)\:=\: i^L z \sum_{i\,=\,1}^2\left(\frac{z+ z_i}{z-z_i}\right)^{L-1},
\end{equation} 
with the conformal map, $\alpha_{\mbox{\tiny $2$}}(z)$ in $\eqref{eq:SLEct}$ corresponding to $L\,=\,2$. Now notice that from $T(z)T(0)\,\sim\,z^{-4}$, we have $\overline{T(z)}\,=\,\bar{z}^{-4}T(\bar{z}^{-1})$, while 
$\overline{\alpha(z)} = -\bar{z}^2 \alpha(\bar{z}^{-1})$. This allows us to re-write the left-hand-side of \eqref{eq:q1} as
\begin{equation}\label{eq:q2}
\sum_i\oint_{\gamma_{z_i}} dz \,\alpha_{\mbox{\tiny $2$}}(z) T(z) \langle\phi^{\mbox{\tiny $(1,s)$}}(z_1)|\mathbb{I}(0)|\phi^{\mbox{\tiny $(1,s)$}}(z_2)\rangle
\end{equation}
where $\gamma_{z_i}$ is a contour encircling the point $z_i$ only. Evaluating \eqref{eq:q2} by mean of the OPE of the stress tensor $T$ with the operators $\mathcal{O}_i$, 
we obtain:
\begin{equation}\label{eq:q7}
\sum_i\oint_{z_i} dz \,\alpha_{\mbox{\tiny $2$}}(z)\,T(z) \langle\phi^{\mbox{\tiny $(1,s)$}}(z_1)|\mathbb{I}(0)|\phi^{\mbox{\tiny $(1,s)$}}(z_2)\rangle\:=\: T_{11}+ T_{12} 
\end{equation}
where 
\begin{equation}
T_{11}\:\equiv\:\sum_i L_0 (z_i)+ 3 z_i L_{-1}(z_i) + 2 z_i^2 L_{-2} (z_i)\quad\mbox{and}\quad
T_{12}\:\equiv\:\,-\,2\, \sum_{j\neq i}\frac{z_i\,z_j}{z^2_{ij}}\,+\,\sum_{j\neq i}z_i \frac{z_i+z_j}{z_i-z_j} \partial_i 
\end{equation}
For $s\,=\,2$, the degenerate boundary primary satisfy the null vector condition \eqref{eq:NVC2}. Thus, using $L_{-1}\phi^{\mbox{\tiny $(1,s)$}}(z_i)\,=\,(\,\partial_i\,-\,\Delta_{\mbox{\tiny $(1,s)$}}z_i^{-1}\,)\,
\phi^{\mbox{\tiny $(1,s)$}}$\footnote{In order to derive this expression, we use the relation between the holomorphic and anti-holomorphic VIrasoro generators arising from the conformal condition on the stress tensor in a BCFT as explicated in \eqref{eq:CBCst} in Section \ref{sec:BCFT} and further, from  the action of rotation on the boundary operator as $\left(z_B L_{-1} - \bar{z}_B \bar{L}_{-1}\right)|\phi^{\mbox{\tiny $(1,2)$}}\rangle\:=\:
\left(z_B \partial_B - \bar{z}_B \bar{\partial}_B\right)|\phi^{\mbox{\tiny $(1,2)$}}\rangle \,$ to obtain the required explicit form for $L_{-1}$.}
we finally resolve eq \eqref{eq:q7} into
\begin{equation}\eqlabel{eq:q8}
 \begin{split}
  &\sum_i\oint_{\gamma_{z_i}} dz \alpha_{\mbox{\tiny $2$}}(z) T(z) \langle\phi^{\mbox{\tiny $(1,s)$}}(z_1)|\mathbb{I}(0)|\phi^{\mbox{\tiny $(1,s)$}}(z_2)\rangle\:=\:
   2\Delta_{\mbox{\tiny{$(1,2)$}}}^2 + 2 ( 1-g) \Delta_{\mbox{\tiny{$(1,2)$}}} + \\
  &\hspace{4cm}+\left(\frac{3\,g}{2}-2 \Delta_{\mbox{\tiny{$(1,2)$}}}-1\right) H_1  +  \sum^2_{i=1} (z_i \partial_i)^2 -\\
  &\hspace{4cm}-2g \Delta_{\mbox{\tiny $(1,2)$}} \sum_{i<j}\frac{z_i z_j}{z^2_{ij}}+ \frac{g}{2} \sum_{i<j} \frac{z_i+z_j}{z_{ij}}(z_i\partial_i-z_j\partial_j),
 \end{split}
\end{equation}
where $H_1\,\equiv\,\sum_{i=1}^2 z_i \partial_i$ is the total momentum of the particles on the unit circle and $z_{ij}\,\equiv\,z_i-z_j$. 
Explicitly noting, 
\begin{equation}
\tilde{H}_2 = \sum_i (z_i \partial_i)^2-4g \Delta_{1,2} \sum_{i<j}\frac{z_i z_j}{z^2_{ij}}+ g \sum_{i<j} \frac{z_i+z_j}{z_{ij}}(z_i\partial_i-z_j\partial_j)\,,
\end{equation} 
to be, up to a similarity transformation, the Calogero-Sutherland (CS) Hamiltonian, while $H_1$ is the first integral invariant of the CS model, we see that 
the boundary correlation function of two level-$2$ degenerate primaries operators $\phi^{\mbox{\tiny $(1,2)$}}$  is an eigenvalue a linear combination of the first two Calogero invariants (See \cite{Estienne:2011qk} for the first 
computation to explicitly demonstrate this).

We wish to analyze AdS$_2$ geodesics or SLE curves whose boundary intersections correspond to higher-level degenerate primaries. The obvious treatment here is to generalize the above discussion to the full L{\"o}wner class map,
{\it i.e.} compute the operation of the conformal map $\alpha_{\mbox{\tiny $L$}}$ for generic $L$. We first decompose the map as 
 \begin{equation}\eqlabel{eq:aLb}
 \alpha_{\mbox{\tiny $L$}}\:=\: i^L \sum_i \sum^{L-1}_{k=0} (2 z_i)^k \left[\frac{1}{(z-z_i)^{k-1}}+ \frac{z_i}{(z-z_i)^k}\right]
 \end{equation}
 we denote the boundary correlator correspondinag to \eqref{eq:aLb} $\langle\phi^{\mbox{\tiny $(1,L)$}}(z_1)|\phi^{\mbox{\tiny $(1,L)$}}(z_2)\rangle$. 
 This leads to the following generalization of \eqref{eq:q7} for arbitrary $L$ as 
 \begin{equation}\eqlabel{q71}
  \begin{split}
  &\sum_{i=1}^2\oint_{\gamma_{z_i}} \alpha_{\mbox{\tiny $L$}} (z) T(z) \langle\phi^{\mbox{\tiny $(1,L)$}}|\phi^{\mbox{\tiny $(1,L)$}}\rangle\:=\: T_{1,L}+ T_{2,L}\:=\\
   &\hspace{1cm}=\:\sum_i \sum^{L-1}_{k=0} {L-1 \choose k} (2 z_i) \left(L_{-k}+ z_i L_{-k-1}\right)\:\\
   &\hspace{1cm}+\sum_{i\neq j}\Delta_i\left[\left(\frac{z_i+z_j}{z_{ij}}\right)^{L-1}+ (L-1) \left(\frac{z_i+z_j}{z_i-z_j}\right)^{L-2}\left(\frac{1}{z_i-z_j}- \frac{z_i+z_j}{z_{ij}^2}\right)\right]\\
   &\hspace{1cm}+\left(\frac{z_i+z_j}{z_i-z_j}\right)^{L-1} z_i \partial_i
  \end{split}
 \end{equation}
 We can further divide  $T_{2,L}$ in the above expression into two groups defined by $L \bmod 2 = 0\,(1)$. For even $L$, we get $T_{2,L}$ to be
 \begin{equation}\eqlabel{eq:q72}
 T_{2,L}\:=\: \Delta_{\mbox{\tiny $(1,L)$}} \sum_{i<j}\left[\left(\frac{z_i+z_j}{z_{ij}}\right)^{L-2}-\left(\frac{z_i+z_j}{z_{ij}}\right)^L\right]\,+\,\sum_{i<j}\left(\frac{z_i+z_j}{z_{ij}}\right)^{L-1}(z_i\partial_i-z_j \partial_j)
 \end{equation}
 while for odd $L$, we have, 
  \begin{equation}\eqlabel{q73}
   T_{2,L}\:=\:\Delta_{\mbox{\tiny $(1,L)$}} \sum_{i<j} 2\left(\frac{z_i+z_j}{z_{ij}}\right)^{L-1}+\sum_{i<j}\left(\frac{z_i+z_j}{z_{ij}}\right)^{L-1}\left(z_i\partial_i+z_j \partial_j\right)
 \end{equation}
We see that $T_{1,L}$ is a combination of Virasoro generators $L_{-m}$ with $0\leq m\leq L$. Taking $\phi^{\mbox{\tiny $(1,L)$}}$ to be the degenerate boundary primary at level $L$ given by the standard equation:
\begin{equation}
\mbox{det}\left\{-J_{-}+ \sum^{L-1}_{m=0} (-g)^m  J^m_{+} L_{-m-1}\right\}|\phi^{\mbox{\tiny $(1,L)$}}\rangle\:=\:0
\end{equation}
where, $J_{-(ij)}= \delta_{ij}-\delta_{i,1}\delta_{j,1}, 1\leq i,j \leq L$ and $J_{+(ij)}= i (l-i) \delta_{i+1,j}$, 
we see that the null vector condition effectively allows a re-writing of $L_{-m}$ terms in $T_{1,L}$ as a polynomial in $L_{-1}$ of degree $L$. This will yield an expression of the form, $\sum_i\sum^L_{j=0} a_j(g)\, z_i^j\,L^j_{-1} $ 
which written in terms of derivative operators, again using, $L_{-1}\phi^{\mbox{\tiny $(1,n)$}}(z_i)\,=\,\left(\,\partial_i\,-\,\Delta_{\mbox{\tiny $(1,2)$}}z_i^{-1}\,\right)\,\phi^{\mbox{\tiny $(1,n)$}}(z_i)$, and  as in the $L=2$ case, yields,
\begin{equation}
T_{1,L}\:=\:\sum_i\sum^L_{j=0} a_j(g) z_i^j \partial_i^j
\end{equation}
Hence:
\begin{equation}\eqlabel{e1}
 \begin{split}
  &\sum_i\left.\oint_{\gamma_{z_i}} \alpha_L(z) T(z) \langle\phi_{1,L}|\phi_{1,L}\rangle\right|_{\mbox{\tiny $\{L \bmod 2=0\}$}}\: =\:\sum_i\sum^L_{j=0} a_j(g) z_i^j \partial_i^j\\
  &\hspace{1cm}+\Delta_{\mbox{\tiny $(1,L)$}} \sum_{i<j}\left[\left(\frac{z_i+z_j}{z_{ij}}\right)^{L-2}-\left(\frac{z_i+z_j}{z_{ij}}\right)^L\right]+\sum_{i<j} (\frac{z_i+z_j}{z_{ij}})^{L-1}\left(z_i\partial_i-z_j \partial_j\right)
 \end{split}
\end{equation} 
while
\begin{equation}\label{e2}
 \begin{split}
  &\sum_i\left.\oint_{\gamma_{z_i}} \alpha_L(z) T(z)\langle\phi^{\mbox{\tiny $(1,L)$}}|\phi^{\mbox{\tiny $(1,L)$}}\rangle\right|_{\mbox{\tiny $L \bmod 2=1$}}\:=\:\sum_i\sum^L_{j=0} a_j(g) z_i^j \partial_i^j\\
  &\hspace{3cm}+\Delta_{1,L} \sum_{i<j} 2\left(\frac{z_i+z_j}{z_{ij}}\right)^{L-1}+\sum_{i<j}\left(\frac{z_i+z_j}{z_{ij}}\right)^{L-1}(z_i\partial_i+z_j \partial_j)
 \end{split}
\end{equation}
The above equations  can be trivially identified with a linear combination of the first $L$ Calogero invariants. 
Hence, to sum up, the SLE curve governed by a $(L-1)$-th order pole L{\"o}wner equation intersects the boundary of the unit disk at points corresponding to boundary operator insertions $|\phi^{\mbox{\tiny $(1,L)$}}\rangle$, and the 
boundary correlator of these two insertions is an eigenfunction of a linear combination of the first $L$ Calogero-Sutherland invariants. The above discussion has been conducted for two particles but can be easily generalized to 
$N\,>\,2$ particles.


\section{BCFT${}_2$ Correlators and the CS Model}\label{sec:BCFT-CS}

Let us now further explore the connection between BCFT${}_2$ and CS model. Specifically, we will discuss the link between the conditions satisfied by correlators containing (degenerate) boundary operators and
the CS integrals of motions. It was already shown that the null vector conditions on correlators of $m$ $\phi^{(1,2)}$ fields and $n$ $\phi^{(2,1)}$ fields can be recast in terms of the CS integrals of motion up
to order $2$ ({\it i.e.} the total momentum $H_1$ and the CS Hamiltonian $H_2^{(g)}$ in \eqref{eq:CSintmot}) \cite{Estienne:2010as}, while a non-vanishing correlators of $N$ degenerate level-$2$ boundary
field and a bulk field placed at the origin is an eigenfunction of the CS Hamiltonian $H_2^{(g)}$ and the conformal dimension of the bulk field can be expressed in terms of the corresponding eigenvalue \cite{Cardy:2003td}.
In this section we review these results and extend them to degenerate operators at higher level and thus to higher order CS invariants. We are interested in the following class of correlators:
\begin{equation}\eqlabel{eq:BCFTcorrs}
 \mathcal{C}_{m+n+1}\left(\left\{z_i\right\},\,\left\{w_j\right\};\,0\right)\:\equiv\:
  \langle\prod_{i=1}^m\phi^{\mbox{\tiny $(r,s)$}}(z_i)\,\prod_{j=1}^n\phi^{\mbox{\tiny $(r',s')$}}(w_j)\,V(0)\rangle,
\end{equation}
where $L\,=\,rs$ and $L'\,=\,r's'$ are the level at which the operators $\phi^{\mbox{\tiny $(r,s)$}}(z_i)$ and $\phi^{\mbox{\tiny $(r',s')$}}(w_j)$ are respectively degenerate. They can be found to be expressed in terms
of eigenfunctions of the CS Hamiltonians via the following class of infinitesimal conformal transformations
\begin{equation}\eqlabel{eq:BBcwiL}
  z\,\longrightarrow\,z+\alpha_{\mbox{\tiny $(\lambda,\lambda')$}}(z),\quad
  \alpha_{\mbox{\tiny $(\lambda,\lambda')$}}(z)\:=\:i^{\lambda}z\sum_{i=1}^m\beta_i\left(\frac{z+z_i}{z-z_i}\right)^{\lambda-1}\hspace{-.1cm}+\:i^{\lambda'}z\sum_{j=1}^n\beta_j\left(\frac{z+w_j}{z-w_j}\right)^{\lambda'-1},
\end{equation}
where $\beta_i$'s are the infinitesimal parameters and, consistently with the notation in \eqref{eq:BCFTcorrs} $(\lambda,\,\lambda')$ are the degeneracy levels of the boundary fields and $z_i$'s and $w_j$'s are their locations 
(with $|z_i|\,=\,1\,=\,|w_j|$). This transformation preserves the interior of the disk (the bulk), satisfying
the relation 
\begin{equation}\eqlabel{eq:BPct}
 \overline{\alpha_{\mbox{\tiny $(\lambda,\lambda')$}}(z)}\:=\:-\overline{z}^2\alpha_{\mbox{\tiny $(\lambda,\lambda')$}}(1/\overline{z}),
\end{equation}
and behaves as a dilatation in a neighbourhood of the origin, while it contains singularities at the insertions of the boundary fields. As a consequence, the conformal Ward identity related to such a transformation receives non-trivial
contributions from these points. More concretely, one way of evaluating the integral 
\begin{equation}\eqlabel{eq:BCFTcwi}
 \begin{split}
  \mathcal{G}_{\mbox{\tiny CWI}}\:=\:&\frac{1}{2\pi i}\int_{\gamma}dz\,\alpha_{\mbox{\tiny $(\lambda,\lambda')$}}(z)
   \langle T(z)\prod_{i=1}^m\phi^{\mbox{\tiny $(r,s)$}}(z_i)\,\prod_{j=1}^n\phi^{\mbox{\tiny $(r',s')$}}(w_j)\,V(0)\rangle-\\
  &\hspace{2cm}-\frac{1}{2\pi i}\int_{\gamma}d\bar{z}\,\overline{\alpha_{\mbox{\tiny $(\lambda,\lambda')$}}(z)}
   \langle\bar{T}(\bar{z})\prod_{i=1}^m\phi^{\mbox{\tiny $(r,s)$}}(z_i)\,\prod_{j=1}^n\phi^{\mbox{\tiny $(r',s')$}}(w_j)\,V(0)\rangle
 \end{split}
\end{equation}
which generates the conformal Ward identities is by deforming the contour of integration closer and closer to the boundary, picking just the contribution of the insertions of the boundary degenerate fields while the integral along the 
rest of the boundary vanishes because of the conformal boundary conditions \cite{Cardy:1984bb}. The conformal boundary conditions themselves connect $\bar{T}(\bar{z})$ inside the disk to the continuation of the holomorphic stress 
tensor outside the disk
\begin{equation}\eqlabel{eq:CBCst2}
 \bar{T}(\bar{z})\:=\:\bar{z}^{-4}T\left(1/\bar{z}\right)
\end{equation}
allowing to compute the contributions from the boundary degenerate fields as a contour integral of just the holomorphic part of \eqref{eq:BCFTcwi} with a clockwise contour which is a direct product of circles around their insertions.
Furthermore, \eqref{eq:CBCst2} relates at the boundary the anti-holomorphic Virasoro generators to the holomorphic ones
\begin{equation}\eqlabel{eq:VirGenRel2}
 \bar{L}_{k}\:=\:(-1)^k\bar{z}_i^{2k}\sum_{l \ge 0}\bar{z}_i^l\frac{(2-k-l)_l}{l!}L_{k+l}
\end{equation}
allowing, together with the infinitesimal conformal transformations $\alpha_{\mbox{\tiny $(\lambda,\lambda')$}}(z)$ ($\lambda\,\le\,L$, $\lambda'\,\le\,L'$), to find  the modes $L_{-k}$ as differential operators acting on
the boundary.

The final result is an eigenvalue equation for the correlator containing degenerate fields showing a linear combination of the Virasoro generators $L_{-k}$ (with $k\,\in\,[0,\,L]$). Then the $L$-th order null-vector condition allows to 
tread the highest order Virasoro generator $L_{-L}$ for a combination of the lower one, with the eigen-value equation which can be written in terms of Calogero-Sutherland commuting operators $H_k$ (up to a similarity transformation), 
with $n$ being at most equal to $L$. Let us see how this works in the cases $L\,=\,2$ and $L\,=\,3$.


\subsection{Bulk-boundary correlators with level-$2$ degenerate primaries }\label{subsec:BBcorr-CS}

For the sake of simplicity and clarity, let us now consider the correlator on the disk of two fields which are degenerate at level $2$, {\it e.g.} two $\phi^{\mbox{\tiny $(1,2)$}}$ and a bulk field $V(w,\bar{w})$. The degenerate fields
are located at the boundary of the disk while the bulk operator at its center. This sub-section reproduces the result firstly obtained in \cite{Cardy:2003td}. 
\begin{equation*}
 \raisebox{-2cm}{\scalebox{.35}{\includegraphics{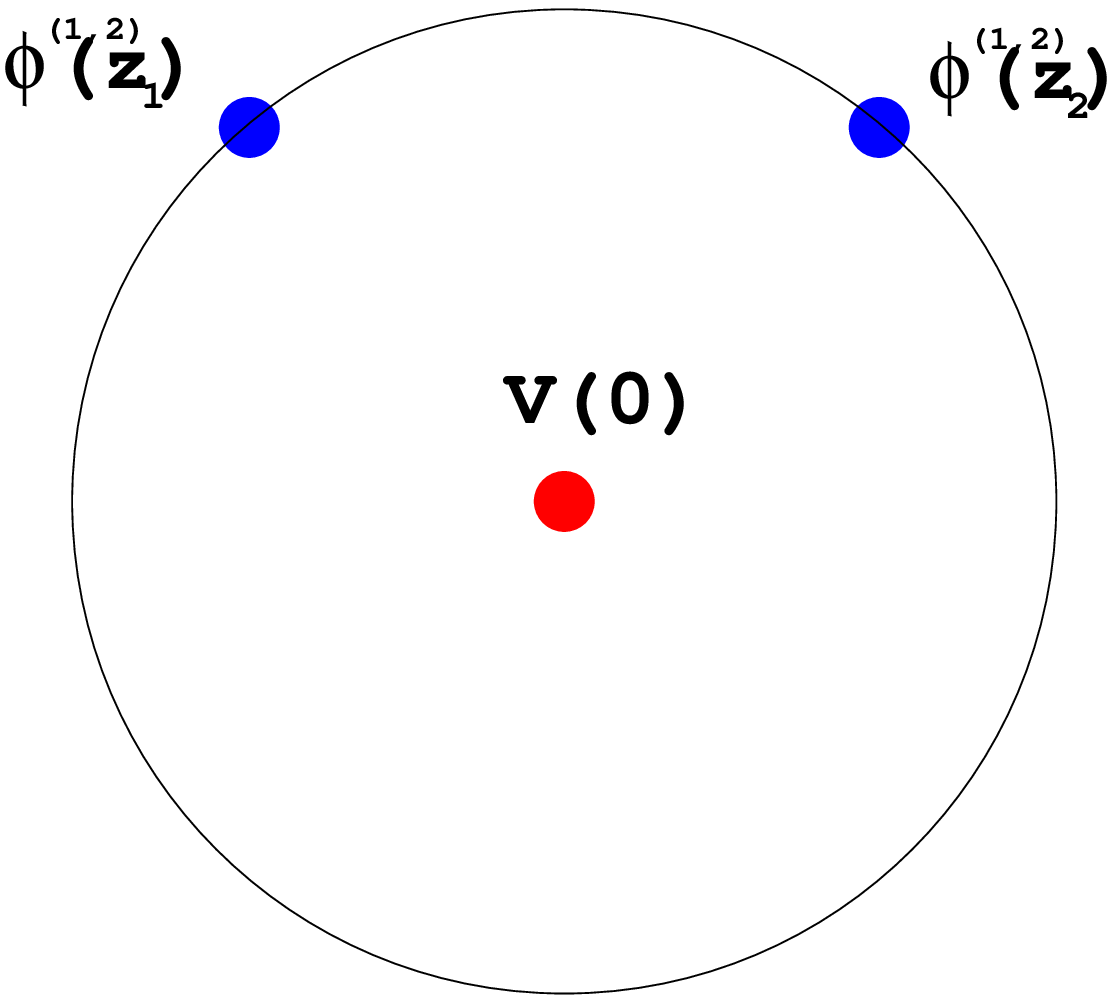}}}
\end{equation*}

Taking the conformal transformation \eqref{eq:BBcwiL} with $L\,=\,2$ (and $m=2$, $n=0$), the Conformal Ward Identity \eqref{eq:BCFTcwi} with a contour $\gamma$ which encircles the center of the disk only returns
\begin{equation}\eqlabel{eq:CWIdisc1}
 \mathcal{G}_{\mbox{\tiny CWI}}\:=\:\left(\sum_{i=1}^2\beta_i\right)\Delta_{\mbox{\tiny V}}\langle\phi^{\mbox{\tiny $(1,2)$}}(z_1)\phi^{\mbox{\tiny $(1,2)$}}(z_2)V(0)\rangle,
\end{equation}
where $\Delta_{\mbox{\tiny V}}$ is the sum of the holomorphic and anti-holomorphic conformal dimentions of $V$. Evaluating now \eqref{eq:BCFTcwi} deforming the contour towards the boundary of the disk as explained in the introductory 
part of this section, using the OPEs for such operators and expanding $\alpha(z)$ around $z_i$'s, the terms of the OPE contributing to the integrals are all from the leading one (containing the 
conformal dimension) to $L_{-2}$. Using now the null vector condition \eqref{eq:NVC2}, one can tread $L_{-2}$ for $g^{-1}L_{-1}^2$, so that $\mathcal{G}_{\mbox{\tiny CWI}}$ acquires the form of the Calogero-Sutherland 
Hamiltonian (up to a similarity transformation) and a constant. Concretely, the integration over the contour of interest yields
\begin{equation}\eqlabel{eq:CWIdisc1b}
 \begin{split}
  \mathcal{G}_{\mbox{\tiny CWI}}\:=\:\frac{2}{g}\beta
   &\left[
    \sum_{i=1}^2\left[z_i^{\Delta_{\mbox{{\tiny $(1,2)$}}}}\left(z_i\partial_{z_i}\right)^2  z_i^{-\Delta_{\mbox{{\tiny $(1,2)$}}}}\right] +
    \frac{g}{2}\left(z_1\partial_{z_1}-z_2\partial_{z_2}\right) - \frac{2g\Delta_{\mbox{\tiny $1,2$}}z_1 z_2}{(z_1-z_2)^2}+g\Delta_{\mbox{\tiny $(1,2)$}}+
   \right.\\
   &\hspace{1cm}\left.+
    \left(\frac{3}{2}g-1\right)\sum_{i=1}^2\left[z_i^{\Delta_{\mbox{{\tiny $(1,2)$}}}}\left(z_i\partial_{z_i}\right) z_i^{-\Delta_{\mbox{{\tiny $(1,2)$}}}}\right]
   \right]
   \langle\phi^{\mbox{\tiny $(1,2)$}}(z_1)\phi^{\mbox{\tiny $(1,2)$}}(z_2)V(0)\rangle,
 \end{split}
\end{equation}
with the first three terms resembling the CS integral of motion at order $2$ -- up to a similarity transformation --, while the last one at order $1$ (notice that the parameters $\beta_i$ have been taken to be equal). 
The similarity transformation which returns a more recognisable form for the second order integral of motion is given by
\begin{equation}\eqlabel{eq:SimTransf}
 \hat{\mathcal{G}}\:\longrightarrow\:|\Psi_2|^{g/2}\hat{\mathcal{G}}|\Psi_2|^{-g/2},\qquad\Psi(z)\:\equiv\:(z_1-z_2)
\end{equation}
with $\hat{\mathcal{G}}$ being the operator in equation \eqref{eq:CWIdisc1b}. Thus, the conformal Ward identity provides an eigenvalue equation for the correlator which involves the second-order Calogero-Sutherland 
Hamiltonian:
\begin{equation}\eqlabel{eq:EigenVeqL2}
 \left[H_{2}^{\mbox{\tiny $(g)$}}-\frac{g^2}{8}-\frac{g-2}{2}\Delta_{\mbox{\tiny $(1,2)$}}\right]\Psi_{2}^{g/2}\langle\phi^{\mbox{\tiny $(1,2)$}}(z_1)\phi^{\mbox{\tiny $(1,2)$}}(z_2)V(0)\rangle\:=\:
 g\Delta_{\mbox{\tiny $V$}}\Psi_2^{g/2}\langle\phi^{\mbox{\tiny $(1,2)$}}(z_1)\phi^{\mbox{\tiny $(1,2)$}}(z_2)V(0)\rangle,
\end{equation}
where the second term on the left-hand-side is proportional to the energy of the ground-state wave-function of $H_2^{\mbox{\tiny $(g)$}}$, while the third one is given by 
$g\left(\frac{\Delta_{\mbox{\tiny $(1,2)$}}}{6}+\frac{c}{12}\right)$,
in agreement with \cite{Cardy:2003td, Doyon:2006ph}.
Joining both equation \eqref{eq:CWIdisc1} and \eqref{eq:EigenVeqL2}, the correlator in question is determined in terms of CS integrals of motion as well as the conformal dimension of the bulk operator is
determined in terms of the eigenvalues of such integrals of motions:
\begin{equation}\eqlabel{eq:CDbulkop}
 \Delta_{\mbox{\tiny V}}\:=\:g^{-1}\left[\Lambda_{2}(g)-\frac{g^2}{8}+\frac{\Delta_{\mbox{\tiny $(1,2)$}}}{6}+\frac{c}{12}\right],
\end{equation}
with $\Lambda_i$ being the eigenvalues of $H_i$. 
As the two boundary 
insertions are brought closer and closer together, the correlator behaves as $|z_1-z_2|^{\sigma}$ with the eigenvalue equation allowing for two solutions according to the OPE of two boundary $(1,2)$-fields
(which contains just the identity operator $\phi^{\mbox{\tiny $(1,1)$}}\,\equiv\,\mathbb{I}$ and $\phi^{\mbox{\tiny $(1,3)$}}$.
Such two solutions correspond to two possible boundary conditions.

Thus, the correlator of two boundary fields which are degenerate at level $2$ and one bulk field is related to an eigenfunction of the Calogero-Sutherland Hamiltonian with eigenvalue $\Lambda_2(g)$. Importantly, this
result is not specific of bulk operators placed at the origin of the disk, and can be mapped for bulk operators at generic point of the disk by the following conformal transformation:
\begin{equation}\eqlabel{eq:BulkCT}
 z\quad\longrightarrow\quad f(z)\:=\:\frac{z+w}{z\bar{w}+1}.
\end{equation}
Notice that this conformal transformation maps boundary points on the disk to boundary point on the disk\footnote{It is straightforward to check that for any boundary point $z_i$, which are such that $z_i\bar{z}_i\,=\,1$, 
$\overline{f(z_i)}\:=\:f^{-1}(z_i)$ and thus $f(z_i)$ still lies on the boundary of the unit disk.}, while the origin in $z$ is now moved to a generic point $w$.

According to eq \eqref{eq:EigenVeqL2}, the correlator of two level-$2$ degenerate primaries $\phi^{\mbox{\tiny $(1,2)$}}$ and a bulk operator $V$, multiplied by the $2$-particle wave-function to the power $g/2$, is an eigenfunction of 
purely the second order CS Hamiltonian $H_2^{\mbox{\tiny $(g)$}}$ and, therefore, it can be expressed in terms of Jack symmetric functions. In \cite{Giombi:2008sr}, this correlator has been related in the semi-classical limit to 
two-fragmented of (global) AdS${}_2$: as $g\,\longrightarrow\,\infty$, the AdS${}_2$ asymptotics survives at both boundaries (thinking of AdS${}_2$ as a BCFT${}_2$ on the strip) and such a behaviour keeps holding at finite $g$. 
Thus, eq \eqref{eq:EigenVeqL2} states that the the two-fragmented AdS${}_2$ solution is determined by the second order CS Hamiltonian and thus it is related the the CS wave-function. Very importantly, the spectrum of the bulk operators
is determined by the spectrum of $H_2^{\mbox{\tiny $(g)$}}$. As we will see in what follows, this statement can be generalised, relating the $s$-fragmented AdS$_2$ solutions to the $s$-th order CS Hamiltonians, and the admitted
boundary conditions are determined by the behaviour of the CS wave-function as a pair of particles are brought closer and closer to each other as prescribed by the relevant eigenvalue equation, which is expressed in terms of 
$H_s^{\mbox{\tiny $(g)$}}$.

\subsection{Bulk-boundary correlators with higher-level degenerate primaries}\label{subsec:BBcorr-CS-L}

Let us now turn to correlators with higher-level degenerate primaries and a bulk operator. For the time being, we focus on the correlators \eqref{eq:BCFTcorrs} with just one type $\phi^{\mbox{\tiny $(r,s)$}}(z_i)$ of boundary
fields. As for the previous discussion, we perform the Conformal Ward Indentity integral \eqref{eq:BCFTcwi} by considering a contour around the origin and then deformining it towards the boundary. The first computation yields
the following expression
\begin{equation}\eqlabel{eq:BBcwiL2}
   \mathcal{G}^{\mbox{\tiny $(L)$}}_{\mbox{\tiny $\beta$}}\:=\:-(-i)^L\left(\sum_{i=1}^m\beta_i\right)\left(\Delta_{\nu}+(-1)^L\bar{\Delta}_{\nu}\right)
   \langle\prod_{i=1}^m\phi^{\mbox{\tiny $(r,s)$}}(z_i)V(0)\rangle,
\end{equation}
where $\Delta_{\nu}$ and $\bar{\Delta}_{\nu}$ are respectively the holomorphic and anti-holomorphic conformal dimensions of the bulk operator $V$. For the sake of concreteness let us specify ourselves to $L\,=\,3$ and choose the
boundary operators to be $\phi^{\mbox{\tiny $(1,3)$}}$. The eigenvalue equation obtained from the Conformal Ward Identity has
the following form
\begin{equation}\eqlabel{eq:CWIeqL3}
 \begin{split}
  &\hspace{3cm}\hat{\mathcal{G}}^{\mbox{\tiny $(3)$}}\mathcal{C}_{\mbox{\tiny $3$}}^{\mbox{\tiny $(1,3)$}}(z_1,\,z_2;\,0)\:=\:-\sigma_{\mbox{\tiny $V$}}\,\mathcal{C}_{\mbox{\tiny $3$}}^{\mbox{\tiny $(1,3)$}}(z_1,\,z_2;\,0),\quad
   \sigma_{\mbox{\tiny $V$}}\:\equiv\:\Delta_{\nu}-\bar{\Delta}_{\nu},\\
  &\hat{\mathcal{G}}^{\mbox{\tiny $(3)$}}\:=\:2\sum_{i=1}^2 z_i^3 L_{-3}^{\mbox{\tiny $(i)$}}+4\sum_{i=1}^2 z_i^2 L_{-2}^{\mbox{\tiny $(i)$}} + 3\sum_{i=1}^2 z_i L_{-1}^{\mbox{\tiny $(i)$}} + 2\Delta_{\mbox{\tiny $(1,3)$}}+\\
  &\hspace{5.5cm}+2\frac{z_1 z_2}{(z_1-z_2)^2}\sum_{i=1}^2 z_i L_{-1}^{\mbox{\tiny $(i)$}} + 4\Delta_{\mbox{\tiny $(1,3)$}}\frac{z_1 z_2}{(z_1-z_2)^2}
 \end{split}
\end{equation}
Using the null vector condition \eqref{eq:NVC3}, after a little algebra\footnote{Notice that the Conformal Ward Identity \eqref{eq:BBcwiL2} for general $\phi^{\mbox{\tiny $(r,s)$}}$ and $L\,=\,1$ tells us that these bulk-boundary
correlators are eigenfunctions of the first order CS Hamiltonian $H_1$ with the eigenvalue being the spin $\sigma_{\mbox{\tiny $V$}}$ of the bulk field.}, one can recast the operator $\hat{\mathcal{G}}^{\mbox{\tiny $(3)$}}$ in the 
following form
\begin{equation}\eqlabel{eq:BBcwiL4}
 \begin{split}
  \hat{\mathcal{G}}^{\mbox{\tiny $(3)$}}\:&=\:\frac{1}{g(1-2g)}
    \left\{
     \sum_{i=1}^2\left(z_i\partial_{z_i}\right)^3+
     \left[
      \Delta_{\mbox{\tiny $(1,3)$}}-\frac{(\Delta_{\mbox{\tiny $(1,3)$}}+1)(\Delta_{\mbox{\tiny $(1,3)$}}+2)}{2}
     \right]
     \sum_{i=1}^2 z_i\partial_{z_i}
    \right.\\
   &\left. -3\Delta_{\mbox{\tiny $(1,3)$}}(\Delta_{\mbox{\tiny $(1,3)$}}+1)\frac{z_1 z_2}{(z_1-z_2)^2}\sum_{i=1}^2 z_i\partial_{z_i}\,+
     \Delta_{\mbox{\tiny $(1,3)$}}(\Delta_{\mbox{\tiny $(1,3)$}}+1)(\Delta_{\mbox{\tiny $V$}}+2)
    \right\}.
 \end{split}
\end{equation}
Notice that, similarly to the level-$2$ case, the operator \eqref{eq:BBcwiL4} can be given in terms of third-order CS Hamiltonian $H_3^{\mbox{\tiny $(g)$}}$. Thus the correlator under examination is related to an eigenfunction of 
$H_3^{\mbox{\tiny $(g)$}}$ and its spectrum determines the conformal dimension allowed for the bulk-field and its spin.

\section{Towards Black Hole Entropy}\label{sec:BH}

We now delineate  the precise physical relevance or meaning of these particles to extremal black hole entropy. In order to clarify this, consider the classic ${\cal N}=4$ IIB 5D extremal $D1-D5-P$ black hole, with a 
U-duality charge invariant, $I = Q_{1} Q_{5} P$, where $Q_1$ and $Q_5$ are the D$1$- and D$5$-brane charges, and $P$ is the momentum on the common direction of the D$1$ and D$5$ branes. The near horizon geometry of this system 
is a BTZ in AdS$_3$ holographically dual to a CFT$_2$ with central charge $c= 6 Q_1 Q_5$. The BTZ black hole is then modeled as a thermal ensemble in one of the chiral 
halves of the CFT$_2$ which lived on the asymptotic circle of the AdS$_3$ background\footnote{This asymptotic circle is nothing but the common direction between the D$1$ and D$5$ branes}. The dilatation eigenvalue of this chiral half is
proportional to the momentum $P$ -- $L_0\,=\,P/(6Q_1 Q_5)$ \cite{Aharony:1999ti}. 
This CFT has a target space which is a $6 Q_1 Q_5$ symmetric product of the compact manifold which form the Neumann directions for the branes. 
As one takes the near-horizon geometry, of this BTZ spacetime, one approaches an AdS$_2 \times S^1$ geometry.  The momentum on the AdS$_3$ circle appears as  charged particles carrying net charge equal to the momentum $P$ in 
AdS$_2$. The trajectory of each of these particles as they emerge from the boundary and are re-absorbed back into it is mappable to a set of interacting symmetric indistinguishable coordinates on the boundary by the SLE transform. 
Hence, counting the  partitions of these  Calogero particles in terms of SLE equivalence classes, will give the degeneracy of this particle bath. As each of these particles is an  excitation in a $6 Q_1 Q_5$ symmetric product field 
theory, it carries $6 Q_1 Q_5$ degrees of freedom and hence, the total number of independent coordinates on the boundary of the unit disk, $N\,=\,6 Q_1 Q_5 P\,=\, 6 I$.  
Given a finite number of particles, $N$, each of these will move along one type of SLE curve corresponding to a particular $L$ and hence to a boundary insertion of $\phi^{\mbox{\tiny $(1,L)$}}$. In principle, one could partition these 
particles into at 
most $N$ classes of SLE curves. Hence, the degeneracy of this particle bath in AdS$_2$is simply equal to the number of partitions of $N$. 
Consequently, the generating function for the dimension of the particle Hilbert space \footnote{We are implictly assuming that these are free particles and operate in a regime where their interaction is negligible. In practical examples 
as in black hole entropy counting this is justified by taking these particles to be BPS.} is 
\begin{equation}
\frac{1}{\Pi_i (1-q^i)}= \sum_N d(N) e^{-\beta N}
\end{equation} where $q=e^{-\beta}$ is a chemical potential for the counting.
This can be inverted to compute the leading degeneracy at large $N$ as 
\begin{equation}\label {3}
d(N) = \int d\beta e^{\beta} \frac{q^{\frac{1}{24}}}{\eta(\beta)}
\end{equation}
where $\eta = \frac{1}{q^{\frac{1}{24}}\Pi_n (1-q^n)}$ is the Dedekind $\eta$-function generating the partition of numbers. 
Using the modular behavior of the $\eta$ function:
\begin{equation}
\eta\left(\frac{4 \pi^2}{\beta}\right)\:=\:\left(\frac{\beta}{2 \pi}\right)^{1/2} \eta (\beta)\,,
\end{equation}
we see that the 'high-temperature' $\beta\rightarrow 0$ limit of 
\eqref{3} yields 
\begin{equation}\label {4}
\left.d(N)\right|_{\beta\rightarrow 0}\:=\:\int d\beta e^{\beta N} \frac{e^{\frac{4 \pi^2}{24 \beta}}}{\eta(\beta)}
\end{equation} 
The saddle point expansion of the RHS of \eqref{4} around $\beta= \sqrt{\frac{\pi^2}{6 N}}$ yields a leading order value given by 
\begin{equation}
d(N)= e^{2 \pi\sqrt\frac{N}{6}}
\end{equation} 
This gives the degeneracy of the system to be 
\begin{equation}
d_{\mbox{\tiny BH}}= d(N)|_{N= 6 I} = 2\pi  \sqrt{I}= 2\pi\sqrt{Q_1 Q_5 P}.
\end{equation}


\section{Summary and Open Questions}\label{sec:Concl}

In this note we modeled the microscopic entropy of a four dimensional static extremal black hole in supergravity in terms of free particles carrying electric charge in the near horizon AdS$_2$ geometry. The primary question here was to 
find a consistent criterion for computing the degeneracy of the Hilbert space of these particles. In other words, we wanted to find a way to classify these particle trajectories. 
We developed this classification by relating the geodesics traced by 
these particles to SLE curves which could be projected onto the boundary of the unit disk. Consequently, they could be encoded in terms of the interactions of the shadow of the curve, under this projection, on the boundary. 
Noticeably,  the L{\"o}wner criterion 
based conformal transformations that performed the 'shadowing' operation on the SLE curve map the region of AdS$_2$,  not in the hull of the curve, to $AdS_2$.  Hence, it  effectively removed the curve from the bulk. The resulting 
interactions governing the shadow boundary points  turn out to be given by linear combinations of the Calogero-Sutherland integral of motion, if the boundary conditions at the points of insertion of the curve with
the unit circle were degenerate boundary operators. Thereby,  in the conformal shadow frame,  where the bulk does not contain a curve, there is  a boundary operator insertion of a primary field of the form 
$|\phi^{\mbox{\tiny $(1,2)$}}\rangle$, at the 
point of intersection of the curve with the boundary. This primary field admitted  a null state at level $2$. In general, under a L{\"o}wner-based shadow transform with an $L$-th order pole, the shadow frame includes the insertions of 
a degenerate boundary operator with a singular vector at level $L$ and which corresponds to \cite{Giombi:2008sr} the  $L$-th order fragmentation of $AdS_2$. The interaction governing the boundary points is a linear combination of the 
first $L$ invariants of the integrable Calogero model.

Thus,  the SLE curves were seen to be divided into equivalence classes based on their ZZ  boundary condition or more directly by  the level $L$ at which the corresponding  singular vector appeared. Therefore, the space of equivalence 
classes was graded by integers. This effectively introduces a quantisation in the space of the SLE curves. 
Hence, counting the degeneracy of this particle bath was equivalent to counting the derangements of the particle number. In the limit of large particle numbers, this was simply the leading expansion of 
the microcanonical ensemble contributing to the reciprocal of the Dedekin $\eta$ function and resulted in an exponential degeneracy. 
In the cases, where this AdS$_2$ is seen as a reduction of the BTZ black hole in AdS$_3$, the particle number is
simply the effective momentum on the asymptotic circle in the AdS$_3$, and this identification yields the Bekenstein-Hawking black hole entropy. 

Note that in the shadow picture, this partition is simply equal to a choice of ZZ boundary 
conditions, and given $N$ particles, the dimension of the Hilbert space is simply the number of ways of distributing these $N$ particles into various boundary operator classes. However, as indicated above, these boundary operators 
corresponds to fragmentations of AdS, in the  conformal shadow frame. Therefore one can potentially think of the black hole microstates as being partitions of AdS$_2$ fragments. 
To sum up, the AdS fragments can be interpreted as microstate geometries for the black hole,  which can be viewed as a manifestation of black hole complementarity. 

Relations between the integrable Calogero model and AdS$_2$ have been explored and studied quite extensively in literature (See \cite{Claus:1998ts, Gibbons:1998fa} for a early interesting approach to analyzing world-line dyamics of 
particles in AdS as Calogero particles as well as \cite{Gaiotto:2004ij} for a counting based on Landau levels of electric particles moving in the magnetically fluxed AdS$_2$ spacetime). However, this note establishes this relation 
clearly and  then uses it to classify the Hilbet space of particles in AdS$_2$, so as to get a handle on the counting problem. 
Further, as one moves from the asymptotic geometry of BTZ in AdS$_3$ to the near-horizon geometry, from the reduced two-dimensional perspective, one moves from a background with a flowing dilaton to that with a fixed one. Hence, this 
probably indicates that AdS$_2$ RG might potentially be modelled  as  a geometric flow in BTZ background. 
There exists a program by Doyon \cite{Doyon:2005wy} to compute CFT correlators simply based on SLE. This should give us a first principle understanding of the 
organisation of field theory states that make a black hole, both from the AdS$_3$ and AdS$_2$ perspective, provided we understand how to map deformations of CFT to those in the SLE language. Lastly, and most importantly, the above 
counting of entropy was accomplished only in cases where the AdS$_2$ was obtained as a reduction of AdS$_3$. An independent first principle derivation of the number of particles, $N$ is still lacking and more study needs to be done on
how the wave functions of the Calogero Hamiltonian capture the SLE curves to arrive at this result. 


\section*{Acknowledgments} 

We would like to thank the organisers of the workshop {\it Stringy geometry} and the MiTP for hospitality and partial support during the workshop, where this project started. S.N gladly acknowledges fruitful and instructive 
conversations with Gary Gibbons, Olaf Lechtenfeld, Gabriel Cardoso and Thomas Mohaupt. P.B. would also like to thank the CAMGSD at the IST Lisbonand Gabriel Cardoso for hospitality while this work was in progress, as well as Kepa Sousa 
for valuable discussions. Finally, P.B. thanks the developers of SAGE \cite{sage}, Maxima \cite{maxima} and Xfig \cite{xfig}. P.B. is supported in part by Plan Nacional de Altas Energ{\'i}as (FPA2015-65480-P) and the Spanish MINECO's 
Centro de Excelencia Severo Ochoa Programme under grant SEV-2012-0249, while S.N. by the FCT fellowship SFRH/BPD/101955/2014.

\appendix

\section{Boundary Correlators of Degenerate Primaries at Level $2$}\label{subsec:Bcorr-CS}

Let us consider correlators involving just boundary fields, which are given in \eqref{eq:BCFTcorrs} with $(r_i,s_i)\,=\,(1,2)$, $(r_j',s_j')\,=\,(2,1)$ (which are the only possible fields which are degenerate at level two)
and no bulk field. In orden to obtain an eigenvalue equation for such correlators in terms just of CS Hamiltonians, one can proceed exactly as in the bulk-boundary correlator case by considering the very same infinitesimal
conformal transformation \eqref{eq:BBcwiL} with $L\,=\,L'\,=\,2$, $\beta_i\,=\,g\,\beta$ and $\beta_j\,=\,\beta$ ($i$ and $j$ run on the $(1,2)$ and $(2,1)$ fields respectively, while $\beta$ is the infinitesimal parameter).
This procedure returns the following eigenvalue equation for $m=2$ $(1,2)$-fiels and $n=2$ $(2,1)$-fields:
\begin{equation}\eqlabel{eq:B4ptcorr}
 \hat{\mathcal{G}}^{\mbox{\tiny $(2)$}}\langle\phi^{\mbox{\tiny $(1,2)$}}(z_1)\phi^{\mbox{\tiny $(1,2)$}}(z_2)\phi^{\mbox{\tiny $(2,1)$}}(w_1)\phi^{\mbox{\tiny $(2,1)$}}(w_2)\rangle\:=\:0
\end{equation}
where the operator $\hat{\mathcal{G}}$ is given by
\begin{equation}\eqlabel{eq:B4ptop}
 \begin{split}
  &\hat{\mathcal{G}}^{\mbox{\tiny $(2)$}}\:\equiv\:
   \sum_{i=1}^2\left(z_i\partial_{z_i}\right)^2+\frac{g}{2}\frac{z_1+z_2}{z_1-z_2}\left(z_1\partial_{z_1}-z_2\partial_{z_2}\right)-2g\Delta_{\mbox{\tiny $(1,2)$}}\frac{z_1z_2}{(z_1-z_2)^2}-\frac{g-2}{2}\Delta_{\mbox{\tiny $(1,2)$}}\,+\\
  &+g\left[
    \sum_{j=1}^2\left(w_j\partial_{w_j}\right)^2+\frac{g^{-1}}{2}\frac{w_1+w_2}{w_1-w_2}\left(w_1\partial_{w_1}-w_2\partial_{w_2}\right)-2\frac{\Delta_{\mbox{\tiny $(2,1)$}}}{g}\frac{w_1w_2}{(w_1-w_2)^2}
     -\frac{g^{-1}-2}{2}\Delta_{\mbox{\tiny $(2,1)$}}
   \right]\,+\\
  &+\sum_{i=1}^2 z_i\partial_{z_i}\,+\,g\sum_{i=j}^2 w_j\partial_{w_j}+\sum_{i,j=1}^2\frac{z_i w_j}{z_i-w_j}\left(\partial_{z_i}-g\partial_{w_j}\right)-
   \left(\Delta_{\mbox{\tiny $(1,2)$}}+g\Delta_{\mbox{\tiny $(2,1)$}}\right)\sum_{i,j=1}^2\frac{z_iw_j}{(z_i-w_j)^2}
 \end{split}
\end{equation}
The equation \eqref{eq:B4ptcorr} can be recast in terms of the CS Hamiltonians via a similarity transformation on the operator $\mathcal{G}^{\mbox{\tiny $(2)$}}$:
\begin{equation}\eqlabel{eq:B4ptop2}
 \hat{\mathcal{G}}^{\mbox{\tiny $(2)$}}\:\longrightarrow\:|\Psi|\hat{\mathcal{G}}^{\mbox{\tiny $(2)$}}|\Psi|^{-1},\qquad 
 \Psi(z,w)\:=\:\left(z_1-z_2\right)^{g/2}\prod_{i,j=1}^2\left(z_i-w_j\right)^{1/2}\left(w_1-w_2\right)^{1/(2g)}
\end{equation}
obtaining that the correlators of just boundary operators which are degenerate at level $2$ are eigenfunctions, modulo the wave-function $\Psi(z,w)$ in \eqref{eq:B4ptop2}, of a linear combinations of the CS Hamiltonians $H_1$ and 
$H_2$ of the $(1,2)$ and $(2,1)$ sectors:
\begin{equation}\eqlabel{eq:B4ptcorrCS}
 0\:=\:
 \left\{
   H_2^{\mbox{\tiny $(g)$}}(z)-H_1(z)+g\left[H_2^{\mbox{\tiny $(1/g)$}}(w)-H_1(w)\right]-\mathcal{E}(g)
 \right\}\Psi_4(z,w)\mathcal{C}_4\left(\{z_i\},\{w_j\}\right),
\end{equation}
wuth $\mathcal{E}(g)$ being a constant dependent on the energy of the ground-state wave-functions of $H_2^{\mbox{\tiny $(g)$}}(z)$ and $H_2^{\mbox{\tiny $(1/g)$}}(w)$ and the central charge $c$. This result was found in a different
way in \cite{Estienne:2010as, Estienne:2011qk}. Notice that the presence of different types of boundary fields -- $\phi^{\mbox{\tiny $(1,2)$}}$ and $\phi^{\mbox{\tiny $(2,1)$}}$ in the current case -- implies that the correlators,
modulo the wave-function $\Psi$, can be expressed in terms of {\it non-polynomial} eigenfunctions of the $H_2^{\mbox{\tiny $(g)$}}$ and $H_2^{\mbox{\tiny $(1/g)$}}$ \cite{Estienne:2011qk}. 

This way of proceeding can be also extended to higher-level degenerate primaries, allowing for a very direct computation of general boundary operators, which are related to the factorisation channels of the bulk-boundary
correlators.


\section{Charged Particles in $AdS_2$ and Calogero Models}
\label{sec:CP-CM}

Let us consider a charged particle propagating in $AdS_2$ with a background gauge field. This setup was analysed in \cite{Maldacena:1998uz}
to study the geodesic trajectories of probe $D0$-branes and in \cite{Pioline:2005pf} to discuss semi-classical Schwinger pair production in relation
to $AdS_2$ fragmentation.  In this appendix, we re-examine it to show that the dynamics of a probe particle in $AdS_2$ with a background constant
electric field is governed by the second order Hamiltonian of Calogero-like models.

For the time being, let us take the following  generic form for the metric
\begin{equation}\eqlabel{eq:genmetric}
 ds^2_2\:=\:g_{\mu\nu}(x)dx^{\mu}dx^{\nu},\qquad
 x^{\mu}\:=\:\left(t,\,\sigma\right),\quad
 A\:=\:-\frac{E}{R^2}\delta^{t}_{\phantom{t}\mu}
        \int^{\sigma}d\sigma'\,\sqrt{-g_{tt}g_{\sigma\sigma}}
  \:\equiv\:\frac{E}{R^2}a(\sigma)\delta^{t}_{\phantom{t}\mu},
\end{equation}
where $g_{\mu\nu}$ is taken to be dependent on the space coordinate $\sigma$ only, $R$ is the radius of $AdS_2$, $A$ is the electric $1$-form potential and $E$ is the constant electric field.
The propagation of such a particle is conveniently described by the worldline action 
\begin{equation}\eqlabel{eq:WLaction}
 S\:=\:\int d\tau\,
  \left\{
   \frac{1}{2}e^{-1}g_{\mu\nu}(\sigma)\dot{x}^{\mu}\dot{x}^{\nu}
   -\frac{1}{2}\,e\,m^2+A_{\mu}\dot{x}^{\mu},
  \right\}
\end{equation}
with $\tau$ being the affine parameter parametrising the worldline
of the particle, the dot $\dot{\phantom{x}}$ indicates the 
derivative with respect to such a parameter, and $e\,=\,e(\tau)$
is the einbein along the worldline, reflecting the invariance of
the action under one-dimensional worldline diffeomorphisms\footnote{For a general look at the worldline formalism see \cite{Strassler:1992zr, Schubert:2001he}. For studies on the particle propagation in curved space-time see
\cite{Bastianelli:2002fv, Bastianelli:2002qw, Bastianelli:2005vk, Bastianelli:2005uy} with a specific analysis of $AdS$/$dS$ backgrounds in \cite{Bastianelli:2008nm}.}.

Let us now substitute the explicit expressions for the metric and
the $1$-form potential \eqref{eq:genmetric} in the action 
\eqref{eq:WLaction}:
\begin{equation}\eqlabel{eq:WLaction2}
 S\:=\:\int d\tau\:
  \left\{
   \frac{1}{2}e^{-1}
   \left[
    g_{tt}(\sigma)\dot{t}^2+g_{\sigma\sigma}(\sigma)\dot{\sigma}^2 
   \right]-
   \frac{1}{2}\,e\,m^2+\frac{E}{R^2}a(\sigma)\dot{t}
  \right\}.
\end{equation}
Notice that this action depends on the time coordinate just through
its derivative and, as consequence, the related momentum is a
first integral of motion on the worldline:
\begin{equation}\eqlabel{eq:pt}
 p_{t}\:=\:e^{-1}g^{tt}(\sigma)\dot{t}+
           \frac{E}{R^2}a(\sigma)\:\equiv\:
 \mbox{const.}
\end{equation}
The existence of such a conserved quantity implies that the model
can be actually reduced to one dimension. It is particularly 
instructive to rewrite the action \eqref{eq:WLaction2} in 
Hamiltonian form:
\begin{equation}\eqlabel{eq:WLaction3}
 \begin{split}
  S\:&=\:\int d\tau\:
   \left\{
    p_{\mu}\dot{x}^{\mu}-e
    \left[
     \frac{1}{2}g^{\mu\nu}(\sigma)
     \left(
      p_{\mu}-\frac{E}{R^2}a(\sigma)\delta^{t}_{\phantom{t}\mu}
     \right)
     \left(
      p_{\nu}-\frac{E}{R^2}a(\sigma)\delta^{t}_{\phantom{t}\nu}
     \right)+
     \frac{1}{2}m^2
    \right]
   \right\}.
 \end{split}
\end{equation}
Notice that the einbein $e$ now behaves as a Lagrange multiplier 
related to the constraint
\begin{equation}\eqlabel{eq:HamConstr}
 \begin{split}
  H\:&=\:\frac{1}{2}g^{\mu\nu}
      \left(
       p_{\mu}-\frac{E}{R^2}a(\sigma)\delta^{t}_{\phantom{t}\mu}
      \right)
      \left(
       p_{\nu}-\frac{E}{R^2}a(\sigma)\delta^{t}_{\phantom{t}\nu}
      \right)+
      \frac{1}{2}m^2\:=\\
     &=\:g^{\sigma\sigma}
      \left[
       \frac{1}{2}p_{\sigma}^2+
       \frac{1}{2}g_{\sigma\sigma}
       \left(
        m^2-(-g^{tt})\frac{E^2}{R^2}a^2(\sigma)
       \right)-
      \frac{p_t\,E}{R^2}g^{tt}g_{\sigma\sigma}a(\sigma)+
      \frac{1}{2}g^{tt}g_{\sigma\sigma}p_t^2
      \right]
      \approx\:0,
 \end{split}
\end{equation}
where the weak equality ``$\approx$'' just indicates that $H$ is 
restricted to be zero but it is not identically zero on the whole 
phase space. 

So far the analysis has been completely classical. At quantum level,
the condition \eqref{eq:HamConstr} selects all the physical states
$|\Psi\rangle$, which have to satisfy it
\begin{equation}\eqlabel{eq:QuantumCond}
 \begin{split}
  H|\Psi\rangle\:&\equiv\:
   g^{\sigma\sigma}\mathcal{H}|\Psi\rangle\:\equiv\\
  &\equiv\:
  g^{\sigma\sigma}
   \left[
     \frac{1}{2}p_{\sigma}^2+
       \frac{1}{2}g_{\sigma\sigma}
       \left(
        m^2-(-g^{tt})\frac{E^2}{R^4}a^2(\sigma)
       \right)-
      \frac{p_t\,E}{R^2}g^{tt}g_{\sigma\sigma}a(\sigma)+
      \frac{1}{2}g^{tt}g_{\sigma\sigma}p_t^2
   \right]|\Psi\rangle\:=\:0,
 \end{split}
\end{equation}
with $\mathcal{H}$ being the term in square brackets.
Actually, one can directly take $\mathcal{H}|\Psi\rangle\:=\:0$
as condition for the physical states. 

Our treatment so far just showed that a charged particle propagating
in a curved background is reducible to a quantum mechanical model
defined by \eqref{eq:QuantumCond}. Let us now specify to some
coordinate patch, starting with the metric in the Bertotti-Robinson 
form:
\begin{equation}\eqlabel{eq:BRmetric}
 ds^2\:=\:-\frac{R^4}{\sigma^4}dt^2+4\frac{R^2}{\sigma^2}d\sigma^2,
     \qquad
 a(\sigma)\:=\:\frac{R^3}{\sigma^2}.
\end{equation}
The physical states are thus defined by
\begin{equation}\eqlabel{eq:QuantumCond2}
 \begin{split}
  0\:&=\:\mathcal{H}|\Psi\rangle\:=\\
     &=\:
      \left[
       \frac{1}{2}p_{\sigma}^2+2\frac{m^2R^2-E^2}{\sigma^2}
       -2\frac{p_t^2}{R^2}\sigma^2+\frac{Ep_t}{R}
      \right]|\Psi\rangle.
 \end{split}
\end{equation}
Notice that this result has been obtained in Lorentz signature. In
Euclidean signature, one obtains a similar expression but with
some sign changed:
\begin{equation}\eqlabel{eq:QuantumCondEucl}
 \begin{split}
  0\:&=\:\mathcal{H}|\Psi\rangle\:=\\
     &=\:
      \left[
       \frac{1}{2}p_{\sigma}^2+2\frac{E^2-m^2R^2}{\sigma^2}
       +2\frac{p_t^2}{R^2}\sigma^2+\frac{Ep_t}{R}
      \right]|\Psi\rangle.
 \end{split}
\end{equation}
Perform the following identifications
\begin{equation}\eqlabel{eq:CMpar}
 \lambda^2\:\equiv\:2\left(E^2-m^2R^2\right),\qquad
 \omega\:\equiv\:2\frac{p_t}{R},
\end{equation}
the condition \eqref{eq:QuantumCond2} resemble the Calogero-Moser
model, with the wave-function satisfying 
\begin{equation}\eqlabel{eq:CMeqWF}
 \left[
  -\frac{1}{2}\left(\frac{d}{d\sigma}\right)^2+
   \frac{\lambda^2}{\sigma^2}+
   \frac{1}{2}\omega^2\sigma^2
 \right]\Psi(\sigma)\:=\:\mathcal{E}\Psi(\sigma),
 \qquad
 \mathcal{E}\:=\:-\frac{p_t E}{R}
\end{equation}
It is important to emphasise that the constant $\mathcal{E}$ is
due to the presence of the constant electric field $E$.

Let us now choose the metric in such a way that it is explicitly
conformally flat, which, to some extent, allows us to treat 
simultaneously the case of Fefferman-Graham coordinates,
global $AdS$ and Schwarzchild coordinates:
\begin{equation}\eqlabel{eq:AFmetric}
 ds^2\:=\:g_{\mu\nu}dx^{\mu}dx^{\nu}\:=\:
        \Omega^2(x)\eta_{\mu\nu}dx^{\mu}dx^{\nu},
 \qquad
 \Omega(x)\:=\:
 \left\{
  \begin{array}{c}
   \frac{R}{\sigma},\qquad\sigma\,\ge\,0\\
   \phantom{R}\\
   \frac{R}{\sin(\sigma)},\quad\sigma\,\in\,[0,\pi]\\
   \phantom{R}\\
   \frac{R}{\sinh(R\sigma)},\quad R\,=\,2\pi\,T_{\mbox{\tiny H}}
  \end{array}
 \right.
\end{equation}
with the function $a(\sigma)$ related to the electric $1$-form is
given by $a(\sigma)\,=\,R^2/\sigma$ in FG coordinates,
$a(\sigma)\,=\,R^2\cot{\sigma}$ in global $AdS$ and
$a(\sigma)\,=\,R^2\coth{R\sigma}$. In these patches
and in Euclidean time, one gets the following conditions on the
physical states:
\begin{equation}\eqlabel{eq:QuantCond3}
 \begin{split}
  &\mbox{Poincar{\'e} coordinates:}\\
  &\qquad0\:=\:\mathcal{H}|\Psi\rangle\:=\:
   \left[
    \frac{1}{2}p_{\sigma}^2+\frac{E^2-m^2R^2}{2\sigma^2}
       +\frac{Ep_t}{\sigma}
       +\frac{p_t^2}{2}
   \right]|\Psi\rangle,\\
  &\mbox{Global coordinates:}\\
  &\qquad0\:=\:\mathcal{H}|\Psi\rangle\:=\:
   \left[
    \frac{1}{2}p_{\sigma}^2+\frac{E^2-m^2R^2}{2\sin^2{\sigma}}
       +Ep_t\,\cot{\sigma}
       +\frac{p_t^2-E^2}{2}
   \right]|\Psi\rangle,\\
  &\mbox{Schwarzchild coordinates:}\\
  &\qquad0\:=\:\mathcal{H}|\Psi\rangle\:=\:
   \left[
    \frac{1}{2}p_{\sigma}^2+
    \frac{E^2-m^2(2\pi\,T_{\mbox{\tiny H}})^2}{2
     \sinh^2{(2\pi\,T_{\mbox{\tiny H}}\sigma})}
    +Ep_t\,\coth{(2\pi\,T_{\mbox{\tiny H}}\sigma)}
       +\frac{p_t^2+E^2}{2}
   \right]|\Psi\rangle.
 \end{split}
\end{equation}
Interestingly, the conditions above represent the Calogero model
(for the Poincar{\'e} patch), the Calogero-Sutherland model and
the Calogero-hyperbolic model, with an extra potential, which is 
Coulomb-like in the first case, $\cot{\sigma}$ in the second one, and
$\coth{\sigma}$ for the last one. Turning the table around,
changing from a Calogero model to the other one is equivalent to
change from an $AdS_2$ coordinate patch to the other one. 

Some comments are in order. First of all, even if it should 
be self-evident, despite the fact that the coordinates have always 
been indicated with $(t,\,\sigma)$, each choice of patch implies, 
in general, a different choice of time and thus a different choice 
of time. Interestingly, the Schwarzchild vacuum 
$|0_{\mbox{\tiny Schw}}\rangle$ is equivalent to the Boulaware one 
$|0_{\mbox{\tiny Bouw}}\rangle$, but inequivalent to the global one 
$|0_{\mbox{\tiny Global}}\rangle$ which is instead equivalent to 
the Poincar{\'e} vacuum $|0_{\mbox{\tiny Poinc}}\rangle$  
\cite{Spradlin:1999bn}. On the other side, the Sutherland model
and hyperbolic one are related by an analytic continuation for
imaginary period of space \cite{Polychronakos:2006nz}.

Secondly, the Calogero-like equations represent the equation
of motion of a scalar in $AdS$. Furthermore, the physical states are
defined at constant $p_t$, which is a first integral of motion, by 
the following eigenvalue (single variable) equation
\begin{equation}\eqlabel{eq:HCalPhys}
 \mathcal{H}_{\mbox{\tiny Cal}}|\Psi\rangle\:=\:
 \mathcal{E}_{\mbox{\tiny Cal}}(p_t,\,E)|\Psi\rangle,
\end{equation}
where $\mathcal{H}_{\mbox{\tiny Cal}}$ and 
$\mathcal{E}_{\mbox{\tiny Cal}}(p_t,\,E)$ are a Calogero 
Hamiltonian operator and the related eigenvalue, respectively.
The eigenvalue $\mathcal{E}_{\mbox{\tiny Cal}}(p_t,\,E)$ depends
on the first integral of motion $p_t$ and on the constant electric
field $E$ and it is given by the Calogero spectrum, establishing
a relation among them and the other Calogero parameters.

Finally, the analysis in \cite{Pioline:2005pf} of the charged particle classical trajectories in $AdS_2$ established that for $E^2-m^2R^2$ negative the particle is confined in the bulk of $AdS_2$, while for  $E^2-m^2R^2$ positive
there are two branches where the particle can be confined (in the case of global $AdS_2$ they correspond to regions close to the two boundaries) and the Schwinger pair production can occur as tunnelling between these two branches.
The case $E^2-m^2R^2=0$ corresponds to BPS particles in a supersymmetric $AdS_2$ background. Identifying the Calogero and $AdS_2$ parameters via the relation $g(g-1)\,\equiv\,m^2R^2-E^2$, the condition for having two branches and
a possible tunnelling process between them can be expressed as $g\,\in[0,\,1]$, {\it i.e.} when the Calogero potential is attractive. Notice that at quantum level $g(g-1)\,\ge\,-1/4$ \cite{Polychronakos:2006nz} and the BF bound becomes 
$m^2R^2\,\ge\,-1/4+E^2$, so that its violation allows for the Schwinger pair production.

\subsection{Global $AdS_2$ and Calogero-Sutherland models}
\label{subsec:GlobCS}

Let us now focus on global $AdS_2$, whose physical states are selected by the Calogero-Sutherland Hamiltonian. More precisely,
because of the presence of the constant electric field in $AdS_2$,
one has an extra term, so that one has a Virasoro $\times\:U(1)$.

Actually, let us start with the Calogero-Sutherland model:
\begin{equation}\eqlabel{eq:HCSglob}
 \left[
  -\frac{1}{2}\partial_{\sigma}+\frac{g(g-1)}{2\sin^2{\sigma}}+
   p_{t}E\cot{\sigma}
 \right]\Psi(\sigma)\:=\:\mathcal{E}\Psi(\sigma),
\end{equation}
which can be identified with $AdS_2$ via the identifications 
$g(g-1)\:=\:m^2R^2-E^2$, $\mathcal{E}\:=\:(p_t^2-E^2)/2$ and 
$\sigma\,=\,[0,\pi]$. In the limits 
$\sigma\,\longrightarrow\,0,\,\pi$, the leading term of the
wave-function is determined by the {\it undeformed} Calogero term.

Let us write the wave function in the form 
$\Psi(\sigma)\:=\:\left(\sin{\sigma}\right)^{\gamma}F(\sigma)$,
so that the Calogero-Sutherland equation becomes
\begin{equation}\eqlabel{eq:HCSglob2}
 \left[
  -\frac{1}{2}\partial_{\sigma}^2-
   \gamma\cot{\sigma}\partial_{\sigma}-
   \frac{\gamma(\gamma-1)-g(g-1)}{2\sin^2{\sigma}}+
   \frac{\gamma^2}{2}+p_t\,E\cot{\sigma}
 \right]F(\sigma)\:=\:\mathcal{E}F(\sigma).
\end{equation}
whose solutions are characterised by $\gamma(\gamma-1)\,=\,g(g-1)$,
{\it i.e.} $\gamma\:=\:g,\,1-g$. A special solution is obtained
by requiring that the term proportional to $\cot{\sigma}$ in the
left-hand-side vanishes, so that equation \eqref{eq:HCSglob2}
can be split into two equation, fixing also the energy 
$\mathcal{E}$ in terms of the Calogero-Sutherland coupling constants 
\begin{equation}\eqlabel{eq:HWsol}
 \begin{split} 
  &\left[-\gamma\partial_{\sigma}+p_t E\cot{\sigma}\right]F(\sigma)
   \:=\:0,\quad\Longrightarrow\quad
   F(\sigma)\:=\:e^{\frac{p_t E}{\gamma}\sigma}\\
  &\left[-\frac{1}{2}\partial_{\sigma}^2+\frac{\gamma^2}{2}\right]
   F(\sigma)\:=\:\mathcal{E}F(\sigma),\quad\Longrightarrow\quad
   \mathcal{E}\:=\:-\frac{1}{2}\left(\frac{p_t E}{\gamma}\right)^2+
   \frac{\gamma^2}{2}
 \end{split}
\end{equation}
Notice that one can recover the value for $\mathcal{E}$ in $AdS_2$
if $\gamma\:=\:-p_t$. However, $\gamma\,=\,g,\,1-g$, so that it
imposes also a relation between $g$, which in terms of $AdS_2$ 
parameters is given below equation \eqref{eq:HCSglob}, and $p_t$:
\begin{equation}\eqlabel{eq:HWpar}
 \gamma\:=\:-p_t\:=\:g\:\quad\Longrightarrow\quad
 -p_{t}\:=\:\frac{1}{2}\left[1\mp\sqrt{m^2R^2-E^2}\right],
 \quad
 1-g\:=\:1+p_{t}\:=\:\frac{1}{2}\left[1\pm\sqrt{m^2R^2-E^2}\right],
\end{equation}
while the solution for $\gamma\,=\,1-g$ is obtained by conjugation
$-p_t\,\longleftrightarrow\,1+p_t$.
This solution corresponds to the highest weight representation in
$CFT_2$ ({\it i.e.} in $AdS_2$), with $-p_t$ being the conformal 
weight of our scalar operator. More general solutions are instead
represented in terms of Jack polynomials.

Notice that, for the solution \eqref{eq:HWsol}, the 
Calogero-Sutherland coupling constant $g$ is -- up to a sign -- 
nothing but the first integral of motion $p_t$, which the eigenvalue of the $L_0$ operator of the (twisted) $SL(2,\mathbb{R})$ of 
$AdS_2$:
\begin{equation}\eqlabel{eq:SL2Rgen}
 \begin{split}
  &\hat{L}_{0}\:=\:i\partial_{t}\:=\:-\hat{p}_t,\\
  &\hat{L}_{\pm 1}\:=\:e^{\pm i\,t}
    \left[
     -\cos{\sigma}\,\hat{p}_t\,\mp i\,\sin{\sigma}\,\hat{p}_{\sigma}
     -\frac{E}{R}\,\sin{\sigma}
    \right].
 \end{split}
\end{equation}
In other words, the Calogero-Sutherland coupling constant $g$ is
related to the highest weight. 

\subsection{Fragmented $AdS_2$ and Calogero models}\label{subsec:FragCal}

Let us now considering a probe charged particle moving in a fragmented $AdS_2$. In order to model it, we consider the following metric
\begin{equation}\eqlabel{eq:FragAdS2soln}
 ds^2_{\mbox{\tiny $2$}}\:=\:\frac{R^2_l}{\sin^2{\left(l\sigma\right)}}\eta_{\mu\nu}dx^{\mu}dx^{\nu},
\end{equation}
where $l\,\in\,\mathbb{Z}_{\mbox{\tiny $+$}}$. It can be thought of as a family of static solutions of Liouville theory \cite{Giombi:2008sr}, with the identification $R_l^2\:=\:l^2/4\pi b^2$  where $b$ is a Liouville parameter.

As for the case of a single $AdS_2$ discussed earlier, we can now study the worldline of the probe particle in this geometry. The computation goes exactly as in the previous case, and the physical states turn out to be selected
by the condition
\begin{equation}\eqlabel{eq:Hl}
 \begin{split}
  0\:&=\:\mathcal{H}_l|\Psi\rangle\:=\\
     &=\:\left[\frac{1}{2}p_{\sigma}^2+\frac{m^2R_l^2-E^2/l^2}{2\sin^2{\left(l\sigma\right)}}+\frac{p_t E}{2l}\cot{l\sigma}-\frac{p_t^2+E^2/l^2}{2}\right]|\Psi\rangle.
 \end{split}
\end{equation}
The worldline Hamiltonian which selects the physical states is again given by a twisted Calogero-Sutherland second-order integral of motion. The number $l$ of the $AdS_2$ fragments become a winding number for the Calogero
circle.

\bibliographystyle{utphys}
\bibliography{gaugegravityrefs}	

\end{document}